\RequirePackage[T1]{fontenc}
\RequirePackage[utf8]{inputenc} 
\RequirePackage[dvipsnames]{xcolor}
\RequirePackage[hyphens]{url}
\documentclass[11pt,a4paper]{article}
\pdfoutput=1
\usepackage{jheppub} 
\usepackage[hyphens]{url} 
\usepackage[dvipsnames]{xcolor}
\usepackage[most]{tcolorbox}

\definecolor{linkc}{rgb}{.8,.15,1}
\usepackage[pagebackref=true,colorlinks=true,linkcolor=linkc,citecolor=linkc,urlcolor=linkc,linktoc=all,pdfusetitle=true]{hyperref}
\renewcommand*{\backref}[1]{}
\renewcommand*{\backrefalt}[4]{{%
		\ifcase #1 
		\or [Cited: pg.~#2.]%
		\else [Cited: pgs. #2.]%
		\fi%
	}}

\usepackage{etoolbox}
\makeatletter
\patchcmd\NAT@citexnum{\let\NAT@last@num\NAT@num}{\MakeLinkTarget[cite]{}\Hy@backout{\@citeb\@extra@b@citeb}\let\NAT@last@num\NAT@num}{}{\fail}
\makeatother

\usepackage[most]{tcolorbox}
\usepackage{amssymb}
\usepackage{amsfonts}
\usepackage{amsbsy}
\usepackage{amsmath}
\usepackage{amsthm}

\usepackage[normalem]{ulem} 
\usepackage{mdframed} 
\usepackage{empheq} 
\usepackage{float}
\usepackage{colortbl} 
\usepackage{multirow} 
\usepackage[bottom,hang,flushmargin]{footmisc}
\usepackage{bm}
\usepackage{accents} 

\DeclareUnicodeCharacter{0393}{\Gamma}
\DeclareUnicodeCharacter{0394}{\Delta}
\DeclareUnicodeCharacter{0398}{\Theta}
\DeclareUnicodeCharacter{039B}{\Lambda}
\DeclareUnicodeCharacter{039E}{\Xi}
\DeclareUnicodeCharacter{03A0}{\Pi}
\DeclareUnicodeCharacter{03A3}{\Sigma}
\DeclareUnicodeCharacter{03A5}{\Upsilon}
\DeclareUnicodeCharacter{03A6}{\Phi}
\DeclareUnicodeCharacter{03A8}{\Psi}
\DeclareUnicodeCharacter{03A9}{\Omega}
\DeclareUnicodeCharacter{03B1}{\alpha}
\DeclareUnicodeCharacter{03B2}{\beta}
\DeclareUnicodeCharacter{03B3}{\gamma}
\DeclareUnicodeCharacter{03B4}{\delta}
\DeclareUnicodeCharacter{03B5}{\epsilon}
\DeclareUnicodeCharacter{03B6}{\zeta}
\DeclareUnicodeCharacter{03B7}{\eta}
\DeclareUnicodeCharacter{03B8}{\theta}
\DeclareUnicodeCharacter{03D1}{\vartheta}
\DeclareUnicodeCharacter{03B9}{\iota}
\DeclareUnicodeCharacter{03BA}{\kappa}
\DeclareUnicodeCharacter{03BB}{\lambda}
\DeclareUnicodeCharacter{03BC}{\mu}
\DeclareUnicodeCharacter{03BD}{\nu}
\DeclareUnicodeCharacter{03BE}{\xi}
\DeclareUnicodeCharacter{03C0}{\pi}
\DeclareUnicodeCharacter{03C1}{\rho}
\DeclareUnicodeCharacter{03C3}{\sigma} 
\DeclareUnicodeCharacter{03C4}{\tau}
\DeclareUnicodeCharacter{03C5}{\upsilon}
\DeclareUnicodeCharacter{03C6}{\phi}
\DeclareUnicodeCharacter{03D5}{\varphi}
\DeclareUnicodeCharacter{03C7}{\chi}
\DeclareUnicodeCharacter{03C8}{\psi}
\DeclareUnicodeCharacter{03C9}{\omega}
\DeclareUnicodeCharacter{21D0}{\Leftarrow}
\DeclareUnicodeCharacter{0212B}{\AA}
\DeclareUnicodeCharacter{00B7}{\cdot}
\DeclareUnicodeCharacter{00B0}{^{\circ}}
\DeclareUnicodeCharacter{266A}{\eighthnote}
\DeclareUnicodeCharacter{266B}{\twonotes}
\DeclareUnicodeCharacter{2202}{\partial}

\def\cA{{\cal A}}

\def\cH{{\cal H}}

\def\cN{{\cal N}}
\def\cO{{\cal O}}

\def\cS{{\cal S}}
\def\cT{{\cal T}}

\def\cV{{\cal V}}

\def\cX{{\cal X}}

\def\IR{{\mathbb{R}}}

\def\IT{{\mathbb{T}}}

\def\IZ{{\mathbb{Z}}}

\def\cA{{\cal A}}

\def\cH{{\cal H}}

\def\cN{{\cal N}}
\def\cO{{\cal O}}

\def\cS{{\cal S}}
\def\cT{{\cal T}}

\def\cV{{\cal V}}

\def\cX{{\cal X}}

\def\bZ{{\mathbb Z}}

\def\sA{{\mathsf A}}
\def\sB{{\mathsf B}}

\def\sD{{\mathsf D}}

\def\sF{{\mathsf F}}

\def\sI{{\mathsf I}}

\def\sN{{\mathsf N}}

\def\sP{{\mathsf P}}

\def\sR{{\mathsf R}}
\def\sS{{\mathsf S}}
\def\sT{{\mathsf T}}
\def\sU{{\mathsf U}}
\def\sV{{\mathsf V}}

\def\sX{{\mathsf X}}

\def\sZ{{\mathsf Z}}

\def\sfa{\mathsf{a}}
\def\sfb{\mathsf{b}}
\def\sfg{\mathsf{g}}

\newcommand{\eqa}[1]{\begin{alignat}{4}#1\end{alignat}}
\newcommand{\beqa}[1]{\begin{empheq}[box=\fbox]{alignat=4}#1\end{empheq}}
\newcommand{\eqas}[1]{\begin{equation}\begin{alignedat}{3}#1\end{alignedat}\end{equation}}
\newcommand{\ov}[1]{\overline{#1}}

\def\hp{\mathsf{h.p.}}
\def\Vir{\mathsf{Vir}}
\newcommand{\orb}{\ensuremath /\!\!/}

\newcommand{\Arf}{\ensuremath \mathsf{Arf}}
\newcommand{\cbos}{\mathsf{cb}}
\newcommand{\wt}{\widetilde}
\newcommand{\sfi}{\mathsf{i}}

\newenvironment{claim}{  \begin{mdframed}[linecolor=black!0,backgroundcolor=black!10]\noindent\itshape\ignorespaces}{\end{mdframed}}

\definecolor{Scolor}{HTML}{ddf1ff}
\definecolor{Tcolor}{HTML}{dfffb7}
\definecolor{Ucolor}{HTML}{fcecf9}
\definecolor{Vcolor}{HTML}{fdfadb}

    \title{A T-Duality of Non-Supersymmetric Heterotic Strings and an implication for Topological Modular Forms}

    \author{Vivek Saxena}
    
    \affiliation{New High Energy Theory Center and Department of Physics and Astronomy,\\ Rutgers University, 126 Frelinghuysen Rd., Piscataway, NJ 08855-0849, USA}
    
    \emailAdd{vivek.hepth@gmail.com}
    
    \abstract{Motivated by recent developments connecting non-supersymmetric heterotic string theory to the theory of Topological Modular Forms (TMF), we show that the worldsheet theory with central charge $(17,\frac{3}{2})$ obtained by fibering the $(E_8)_1 \times (E_8)_1$ current algebra over the $\mathcal{N}=(0,1)$ sigma model on $S^{1}$ with antiperiodic spin structure (such that the two $E_8$ factors are exchanged as we go around the circle), is continuously connected to the $(E_8)_2$ theory in the Gaiotto$-$Johnson-Freyd$-$Witten sense of going ``up and down the RG trajectories''. Combined with the work of Tachikawa and Yamashita, this furnishes a physical derivation of the fact that the $(E_8)_2$ theory corresponds to the unique nontrivial torsion element $[(E_8)_2]$ of $\mathsf{TMF}^{31}$ with zero mod-2 elliptic genus.
    }

\begin{document}
    \setcounter{tocdepth}{2}
    \maketitle

\newpage
\section{Introduction}\label{sec:intro}
Given a quantum field theory $\cT$ parametrized by some set of generalized couplings or deformations $\{\lambda^i\}$, it is a natural question to ask whether the set of such $\cT(\{\lambda^i\})$ admits any interesting topology. Indeed, this is necessary for any attempt to rigorize the notion of  ``the space $\cS(\cT)$ of theories,'' and investigate, for instance, the set of path-connected components $\pi_{0}(\cS(\cT))$ or connected components $H^{0}(\cS(\cT))$. 

Recently, Gaiotto, Johnson-Freyd, and Witten introduced the notion of ``flowing up and down the RG trajectories'' \cite{Gaiotto:2019asa}. Given a 2d $\cN =(0,1)$ supersymmetric quantum field theory (SQFT) $\cT$, one introduces a supersymmetric mass deformation to yield a theory $\cT'$, which is equivalent to $\cT$ at long distances. Then one supersymmetrically perturbs $\cT'$ to yield a third theory $\cT''$. These authors were interested in investigating whether the perturbation to $\cT''$ can trigger spontaneous supersymmetry breaking. They effectively described a sequence of theories $\cT \to \cT' \to \cT''$ that are continuously connected in a family of supersymmetric theories. With these supersymmetric deformations, the theories $\cT$, $\cT'$, and $\cT''$ are in the same ``homotopy class.''

There is an interesting, albeit conjectured, connection between 2d $\mathcal{N}=(0,1)$ SQFTs and the generalized cohomology theory of Topological Modular Forms (TMF) \cite{Hopkins95,Hopkins2002,Hopkins:2002rd,Goerss:2009,Lurie2009,Douglas2014-cl,BrunerRognes}. Building on some earlier work by Segal \cite{Segal88,Segal2007}, Stolz and Teichner \cite{StolzTeichner1,StolzTeichner2} conjectured that to every 2d $\mathcal{N} = (0,1)$ SQFT, there corresponds a class in TMF. This map is expected to descend to homotopy classes of 2d $\mathcal{N}=(0,1)$ SQFTs, which are naturally graded by an anomaly coefficient $\nu \in \IZ$, leading to a map\footnote{We define $\mathsf{TMF}^{-\nu} := \mathsf{TMF}^{-\nu}(\mathrm{pt})$, where $\mathrm{pt}$ denotes a point. The conjecture stated here is a special case (for spin SCFTs) of some wide-ranging results; see \cite{Gukov:2018iiq,Tachikawa:2021mby} for detailed expositions, including the corresponding equivariant and twisted versions.} 
\begin{align}
  \mathsf{DefClass} : \left\{ \begin{array}{c} \text{homotopy class of }\\ \text{2d $\cN=(0,1)$ SQFT with anomaly $\nu$}\end{array}\right\} &\longmapsto \mathsf{TMF}^{-\nu} ~. \label{eq:defclass}
\end{align}  
 Assuming the validity of the Segal-Stolz-Teichner conjecture, TMF classes make it possible to study, among other things, the path-connected components of the space of 2d $\cN=(0,1)$ theories. (We will review some aspects of this in Section \ref{sec:tmf}.) For string theorists, 2d $\cN=(0,1)$ superconformal field theories (SCFTs) arise rather naturally as worldsheet CFTs for the heterotic string \cite{Gross1,Gross2,Gross3}.\footnote{For a 2d $\cN=(0,1)$ SCFT with central charge $(c_L, c_R)$, the anomaly coefficient is $\nu := -2(c_L - c_R)$.} A detailed investigation under the assumption of the Segal-Stolz-Teichner conjecture \eqref{eq:defclass} led Tachikawa and Yamashita \cite{Tachikawa:2021mvw,Tachikawa:2021mby} to conclude that there are no global anomalies in the heterotic string. This requires a careful analysis of the subtle torsion invariants measured by TMF classes.\footnote{See also \cite{Gaiotto:2018ypj,Gukov:2018iiq,Gaiotto:2019gef,Johnson-Freyd:2020itv,Lin:2021bcp,Yonekura:2022reu,Lin:2022wpx,Albert:2022gcs,BerwickEvans,Debray:2023yrs,Debray:2023rlx,Johnson-Freyd:2024rxr,Tachikawa:2024ucm} for some other recent discussions of TMF.} 

The perturbative heterotic string in the NSR formalism in lightcone gauge \cite{Polchinski:1998rr,Blumenhagen:2013fgp} takes as input a modular-invariant 2d $\mathcal{N}=(0,1)$ worldsheet SCFT with central charge $(c_{\mathsf{left}}, c_{\mathsf{right}}) = (24, 12)$, and yields as outputs the spacetime Hilbert space (spacetime spectrum) and the spacetime scattering amplitudes. Computing scattering amplitudes involves an integration over the moduli space of super-Riemann surfaces. In the simplest case, when the Riemann surface has the topology of a 2-torus $\mathbb{T}^2$, this integration involves a sum over the four spin structures on $\mathbb{T}^2$, which amounts to a GSO projection \cite{Gliozzi:1976jf,Gliozzi:1976qd,Seiberg:1986by}. 

This work will restrict our attention to 2d $\cN = (0,1)$ SCFTs that model the internal worldsheet degrees of freedom of two nine-dimensional non-supersymmetric heterotic string theories. Non-supersymmetric string theories have been known for years \cite{Alvarez-Gaume:1986ghj,Dixon:1986iz,Seiberg:1986by,Kawai:1986vd,Ginsparg:1986wr,Itoyama:1986ei,Forgacs:1988iw,Bergman:1997rf,Blum:1997gw,Mikhailov:1998si,Bergman:1999km,Martinec:2009ks} and their heterotic versions have since been revisited in different contexts \cite{Suyama:2001bn,Horava:2007yh,Horava:2007hg,Hellerman:2007zz,Ashfaque:2015vta,Blaszczyk:2015zta,GrootNibbelink:2016pta,Faraggi:2019drl,Faraggi:2020wld,Kaidi:2020jla,BoyleSmith:2023xkd,Kaidi:2023tqo,Nakajima:2023zsh,Avalos:2023ldc,Fraiman:2023cpa,Basile:2023knk,DeFreitas:2024ztt}. The non-supersymmetric heterotic strings we consider here have a spacetime tachyon. Nevertheless, their worldsheet theories are well-defined modular-invariant 2d $\cN=(0,1)$ SCFTs. 

For a general $d$-dimensional heterotic string model (where $2 < d \leq 10$), the internal degrees of freedom correspond to a modular-invariant 2d CFT with $(c_{\mathsf{left}}, c_{\mathsf{right}})_{\mathsf{internal}} = (26-d, \frac{3}{2}(10-d))$.\footnote{The spacetime degrees of freedom contribute $(c_{\mathsf{left}}, c_{\mathsf{right}})_{\mathsf{spacetime}} = (d-2, \frac{3}{2}(d-2))$.} 
 It is a general fact that internal bosonic (resp. fermionic) CFTs\footnote{Fermionic (or spin) CFTs require a choice of spin structure for their definition, whereas bosonic CFTs do not.} lead to spacetime supersymmetric (resp. spacetime non-supersymmetric) theories \cite{BoyleSmith:2023xkd}. The deformation class of an internal 2d $\cN=(0,1)$ SCFT for the $d$-dimensional non-supersymmetric heterotic string thus determines a class $[\sT] \in \mathsf{TMF}^{22+d}$ by \eqref{eq:defclass}. A key motivation for this work is the following question:
\begin{claim}
  Given two $2d$ $\mathcal{N}=(0,1)$ SQFTs $\cT$ and $\cT'$ with anomaly coefficient $\nu$, are they in the same homotopy class? Equivalently, do they determine the same class in $\mathsf{TMF}^{-\nu}$ in the image of \eqref{eq:defclass}?
\end{claim}
In particular, if the image of the forward map \eqref{eq:defclass} is mathematically intractable to compute for $\cT'$, one can ask if it is possible to find another 2d $\mathcal{N}=(0,1)$ theory $\cT$ in the same homotopy class, for which the image \eqref{eq:defclass} is computable using, for example, the results of \cite{Tachikawa:2021mby,Tachikawa:2023nne,Tachikawa:2023lwf}. Here, we will consider the following $d = 9$ spacetime non-supersymmetric heterotic theories:
  \begin{itemize}
    \item Theory $1$: the model obtained by fibering the $(E_8)_1 \times (E_8)_1$ current algebra over the $\cN=(0,1)$ $\sigma$-model on $S^1$ with antiperiodic spin structure, such that the two $E_8$ factors are exchanged as one goes around  the circle.

    \item Theory $2$: the $(E_8)_2$ current algebra with a single right-moving Majorana-Weyl fermion on $S^1$ with antiperiodic spin structure, with a $\mathbb{Z}_2$ action.
  \end{itemize}
The $\cN = (0,1)$ $\sigma$-model on a circle of radius\footnote{We use the notation $S^{1}_{\mathsf{R}} := \IR/(2\pi\mathsf{R}\IZ)$ for a circle of radius $\mathsf{R}$.} $R$ consists of a compact boson $X$ (which we denote by $\cbos_{R}$) along with its superpartner, a right-moving Majorana-Weyl fermion, denoted by $\wt{\psi}$. The internal worldsheet $\cN = (0,1)$ theories with central charge $(c_L, c_R) = (17,\frac{3}{2})$ corresponding to the above spacetime theories are
  \begin{itemize}
    \item $\sT_1 \otimes \wt{\psi}$ : the tensor product of a fermionization of the product theory $\cbos_{2R} \otimes (E_8)_1 \times (E_8)_1$, with a right-moving Majorana-Weyl fermion $\wt{\psi}$. More precisely,
    \eqa{
    &\sT_1 &&:= \left[\cbos_{2R} \otimes (E_8)_1 \times (E_8)_1 \otimes \Arf\right]\orb\mathsf{diag}\big( \IZ_2^\hp \times \IZ_2^\sigma \times \IZ_2^\Arf \big) ~,
    }
    where the $\Arf$ theory \cite{Debray2018,Karch:2019lnn,Hsieh:2020uwb} is a fermionic invertible phase \cite{Freed:2004yc} with a $\IZ_2^\Arf$ symmetry, $\IZ_2^\hp$ is the half-period shift symmetry of the compact boson (under which $X \mapsto X +  2\pi R$), and $\IZ_2^\sigma$ flips the two $(E_8)_1$ factors.\footnote{The fermionization of the compact boson with the $\Arf$ theory crucially endows the spacetime fermions a spacetime antiperiodic spin structure around $S^{1}_{R}$, see Section \ref{sec:generaltduality}.} The theory $\sT_1 \otimes \wt{\psi}$ is the worldsheet theory for the angular part of the 7-brane of \cite{Kaidi:2023tqo}. 
    \item $\sT_2 \otimes \wt{\psi}$ : the tensor product of a fermionization of the product $\cbos_{2R'}
    \otimes (E_8)_2 \times \lambda$ with a right-moving Majorana-Weyl fermion $\wt{\psi}$. More precisely,
    \eqa{
    &\sT_2 &&:= \left[\cbos_{2R'} \otimes (E_8)_2 \times \lambda\right]\orb\mathsf{diag}\big( \IZ_2^\hp \times (-1)^{F} \big) ~,
    }
    where $(E_8)_2 \times \lambda$ is a fermionic theory obtained from $(E_8)_1 \times (E_8)_1$ by fermionization:
    \eqa{
       &(E_8)_2 \times \lambda &&:= \big((E_8)_1 \times (E_8)_1 \otimes \Arf\big)\orb\mathsf{diag}\big( \IZ_2^\hp \times \IZ_2^\Arf \big) ~,
    }
    and comes equipped with a fermion parity symmetry generated by $(-1)^{F}$. 
  \end{itemize} 
  
  In this work, we show that\footnote{The T-duality of $\sT_1$ and $\sT_2$ morally follows from much earlier work of Ginsparg and Vafa \cite{Ginsparg:1986wr}. See also \cite{Nakajima:2023zsh}.}
  \begin{tcolorbox}[bicolor, colback=blue!15,
    colbacklower=white, boxrule=0pt,frame hidden]
  \begin{enumerate}
    \item Given a bosonic 2d CFT $\cT_\sA$ with a non-anomalous $\IZ_2^{(\sA)}$ symmetry, the 2d spin CFTs
    \eqa{
      \cX_1 &:= \big(\cbos_{2R} \otimes \cT_{\sA} \otimes \Arf)\orb\mathsf{diag}\big( \IZ_2^\hp \times \IZ_2^{(\sA)} \times \IZ_2^\Arf\big) ~,
    }
    and
    \eqa{
     \cX_2 := \big(\cbos_{2R'} \otimes \cT_{\sF'} \big)\orb \mathsf{diag}\big( \IZ_2^\hp \times (-1)^{F}\big) ~,
    }
    are T-dual for $R' = \frac{1}{2R}$. Here $\cT_{\sF'}$ is the tensor product of the fermionization of $\cT_\sA$ with the $\Arf$ theory, i.e., $\cT_{\sF'} := (\cT_{\sA} \otimes \Arf)\orb \mathsf{diag}(\IZ_2^{(\sA)}\times \IZ_2^\Arf) \otimes \Arf$. This T-duality follows from that of the compact boson, and the theory $\cT_\sA$ serves merely as a spectator in this duality.
    
    \item $\mathsf{T}_{1}$ is T-dual to $\mathsf{T}_{2}$ for $R'=\frac{1}{2R}$.  This is a special case of the previous result when $\cT_\sA$ is taken to be the bosonic holomorphic $(E_8)_1 \times (E_8)_1$ theory. 
    
    Therefore, the two spacetime non-supersymmetric heterotic string theories with internal 2d $\cN=(0,1)$ SCFTs $\sT_{1} \otimes \wt{\psi}$ and $\sT_{2} \otimes \wt{\psi}$ are T-dual to each other.
    \item The worldsheet theory $\sT_{1} \otimes \wt{\psi}$ is continuously connected to the 2d $\cN=(0,1)$ $(E_8)_{2}$ SQFT obtained by a certain tachyonic deformation of $\sT_{2} \otimes \wt{\psi}$ that leaves only the $(E_8)_2$ current algebra as the light degrees of freedom -- we often refer to this as the ``$(E_8)_2$ theory''. Therefore, the $\sT_{1} \otimes \wt{\psi}$ theory and the $(E_8)_2$ theory are related by moving up and down the RG trajectories.
  \end{enumerate}
\end{tcolorbox}
Combining the above with the results of \cite{Tachikawa:2021mby,Tachikawa:2023nne}, we give a physical derivation of the following fact (cf. Conjecture A.10 of \cite{Tachikawa:2023lwf}):
\begin{tcolorbox}[bicolor, colback=blue!15,
  colbacklower=white, boxrule=0pt,frame hidden]
    The $(E_8)_{2}$ theory corresponds to the unique nontrivial torsion element $[(E_8)_2]$ in $\mathsf{TMF}^{31}$ with zero mod-$2$ elliptic genus.
\end{tcolorbox}
This paper is organized as follows: in Section \ref{sec:modernz2orbifoldintro}, we review $\bZ_2$ orbifolds of 2d CFTs as a preparation for later sections. A key ingredient here is the fermionization of a bosonic theory by the $\mathsf{Arf}$ theory mentioned above. In Section \ref{sec:fermionizedtheories}, we use fermionization to construct building blocks leading up to theories $\sT_1$ and $\sT_2$. In Section \ref{sec:generaltduality}, we discuss a general T-duality between two fermionic CFTs, a special case of which leads to the T-duality of the theories $\sT_{1}$ and $\sT_{2}$. Along the way, we also comment on the effect of worldsheet fermionization on \textit{spacetime} spin structure. In Section \ref{sec:tachyon}, we explain how a tachyonic deformation added to theory $\sT_{2} \otimes \wt{\psi}$ induces an RG flow to the $(E_8)_{2}$ theory, thereby proving that the two theories are continuously connected in the space of theories. Finally, in Section \ref{sec:tmf}, we discuss a consequence of these findings for TMF, and in Section \ref{sec:conclusions}, we list our conclusions and outline some future directions.

In Appendix \ref{app:ThetaFunc}, we review the Jacobi theta functions and their $SL(2,\IZ)$ transformations. In Appendix \ref{app:compchar}, we review the characters of $(E_8)_1 \times (E_8)_1$ and $(E_8)_2$, and in Appendix \ref{app:factoringintcft}, we review the factoring out of universal (super)Virasoro contributions in torus partition functions. 

Readers familiar with these worldsheet models could skip directly to Sections \ref{sec:generaltduality}, \ref{sec:tachyon}, and \ref{sec:tmf}.

\acknowledgments 
I thank Yuji Tachikawa and Kazuya Yonekura for suggesting this investigation and for various discussions. I especially thank Yuji Tachikawa for a careful reading of an earlier version of the draft and for explaining various details of his work on TMF. I also thank Tom Banks, Zohar Komargodski, Sergei Lukyanov, Gregory Moore, and Martin Ro\v{c}ek for discussions, and the referee for helpful comments and suggestions.

This work was supported by the US Department of Energy under grant DE-SC0010008.

\section{Review of $\IZ_2$ orbifolds of 2d CFTs}
\label{sec:modernz2orbifoldintro}
This section is a self-contained pedagogical review of the modern perspective on $\bZ_2$ orbifolds of 2d CFTs as originally developed in \cite{Hsieh:2020uwb} -- readers familiar with this work may safely skip this section. 

Consider a $(1+1)d$ \underline{bosonic} theory $\sA$ with a non-anomalous $\bZ_2$ symmetry denoted by $\bZ_2^{(\sA)}$. On a spatial $S^1$, the Hilbert space of $\sA$ can be twisted or untwisted, depending on whether or not one introduces a twist by the $\bZ_2^{(\sA)}$ symmetry around the spatial circle. We decompose the untwisted Hilbert space of $\sA$ in terms of sectors that are even (resp. odd) under the $\bZ_2^{(\sA)}$ action, namely $\cH_{\sA,\sS}$ (resp. $\cH_{\sA,\sT}$): $\cH_{\sA}^{\mathsf{untwisted}} = \cH_{\sA,\sS} \oplus \cH_{\sA,\sT}$. The $\bZ_2^{(\sA)}$-twisted Hilbert space of $\sA$ can likewise be decomposed into even (resp. odd) sectors, denoted by $\cH_{\sA,\sU}$ (resp. $\cH_{\sA,\sV}$): $\cH_{\sA}^{\mathsf{twisted}} = \cH_{\sA,\sU} \oplus \cH_{\sA,\sV}$.  The partition function of the theory $\sA$ is then, quite simply,
\eqa{
    \sZ_{\sA} &= \sZ_{\sS} + \sZ_{\sT} ~.
}
Orbifolding the theory $\sA$ by $\IZ_2^{(\sA)}$ yields the bosonic theory $\sD := \sA\orb\bZ_2^{(\sA)}$, the partition function of which is
\eqa{
    \sZ_{\sD} &= \sZ_{\sS} + \sZ_{\sU} ~.
}
The theory $\sD$ has a $\bZ_{2}^{(\sD)}$ symmetry and orbifolding by it gives back theory $\sA$ \cite{Vafa:1989ih}. In the $\bZ_{2}^{(\sD)}$-twisted Hilbert space, the $\bZ_{2}^{(\sD)}$- even and odd sectors are $\sT$ and $\sV$ respectively. We list the states of $\sA$ and $\sD$ in Table \ref{tbl:stateAandD}.
\begin{table}[H]
    \centering
    \scalebox{0.95}{%
    \begin{tabular}{|c|c|c|} \cline{2-3}
        \multicolumn{1}{c|}{$\sA$} & untwisted & twisted \\ \hline
        even & \cellcolor{Scolor}{$\sS$} & \cellcolor{Ucolor}{$\sU$} \\ \hline
        odd & \cellcolor{Tcolor}{$\sT$} & \cellcolor{Vcolor}{$\sV$} \\ \hline
    \end{tabular}
    \quad
    \begin{tabular}{|c|c|c|} \cline{2-3}
        \multicolumn{1}{c|}{$\sD$} & untwisted & twisted \\ \hline
        even & \cellcolor{Scolor}{$\sS$} & \cellcolor{Tcolor}{$\sT$} \\ \hline
        odd & \cellcolor{Ucolor}{$\sU$} & \cellcolor{Vcolor}{$\sV$} \\ \hline
    \end{tabular}
    }
    \caption{\label{tbl:stateAandD}States of theories $\sA$ and $\sD$.}
\end{table}
One can fermionize the theory $\sA$ by tensoring it with a fermionic theory called the $\Arf$ theory, whose lowest energy state on the spatial circle is nondegenerate.\footnote{The $\Arf$ theory is an invertible topological theory \cite{Freed:2004yc}, with a one-dimensional Hilbert space. It can be understood as the extreme low-energy limit of the nontrivial topological phase of the Kitaev chain \cite{Kitaev:2000nmw} or the infrared limit of a mass deformation of the $(c_L, c_R) = (\frac{1}{2}, \frac{1}{2})$ Ising CFT. See \cite{Karch:2019lnn,Kaidi:2019tyf,Kaidi:2019pzj} for more details.} As a fermionic theory, it has a $\bZ_{2}$ (fermion number) symmetry denoted by $\IZ_2^\Arf$, with eigenvalues depending on the periodicity of the ground state. The symmetry is generated by the fermion number operator\footnote{\label{foot:arfinvt}For a Riemann surface $\Sigma$ with spin structure $\rho$, $\sF_\Arf := \text{Arf}[\Sigma,\rho]$ is the Arf invariant of a quadratic refinement of the intersection form on $H_{1}(\Sigma, \bZ_2)$ \cite{Johnson1980}. In particular, when $\Sigma = \mathbb{T}^2$ is the 2-torus, $\text{Arf}[\Sigma, \rho] = 1$ if $\rho = (\sR,\sR)$ and $\text{Arf}[\Sigma, \rho] = 0$ if $\rho = (\sN\sS,\sN\sS), (\sN\sS, \sR)$ or $(\sR, \sN\sS)$, where $\sN\sS$ (antiperiodic) and $\sR$ (periodic) denote the spin structures around the two homologically nontrivial $1$-cycles \cite{Alvarez-Gaume:1986rcs}. See \cite{Debray2018} for a nice pedagogical review of the $\Arf$ invariant and the $\Arf$ theory.}
\eqa{
    \left.(-1)^{\sF_{\Arf}}\right|_{\text{ground state}} &= \left\{
      \begin{array}{ll}
        +1 & \text{antiperiodic ($\sN\sS$)} ~,\\
        -1 & \text{periodic ($\sR$)} ~.
      \end{array}
    \right. \label{eq:fermkitaev}
}
The product theory $\sA \otimes \Arf$ has a $\bZ_2^{(\sA)} \times \bZ_2^{\Arf}$ symmetry. One can define a \underline{fermionic} theory $\sF$ from it by orbifolding by the diagonal $\IZ_2$:
\eqa{
    &\sF := (\sA \otimes \Arf)\orb \mathsf{diag}(\bZ_2^{(\sA)} \times \bZ_2^{\Arf}) ~.
} 
The states of $\sF$ are states in the twisted and untwisted Hilbert spaces of $\sA \otimes \Arf$ that are even under the diagonal $\bZ_2$. Let $Q_{\sA}$ denote the $\bZ_{2}^{(\sA)}$ charge. States that survive the orbifold satisfy $Q_{\sA}(-1)^{\sF_{\Arf}} = +1$. We decompose the four sectors of $\sA \otimes \Arf$ (untwisted or twisted with respect to each $\bZ_2$ factor) in terms of two labels $(u_{\sA}, s_{\sF})$ where $u_{\sA}=+1$ (resp. $u_{\sA}=-1$) denotes untwisted (resp. twisted) with respect to the $\bZ_2^{(\sA)}$ action, and $s_{\sF}$ denotes the fermion periodicity of $\sF$ which is either $s_{\sF} = +1$ (periodic ($\sP$)) or $s_{\sF}=-1$ (antiperiodic ($\sA\sP$)). Now, \eqref{eq:fermkitaev} implies that $(-1)^{\sF_{\Arf}} = -s_{\sF}u_{\sA}$. So, the states of $\sF$ satisfy $Q_{\sA} = -s_{\sF}u_{\sA}$, and are called bosonic (resp. fermionic) if they come from states with $(-1)^{\sF_{\Arf}} = +1$ (resp. $-1$), see Table \ref{tbl:stateF}. 
\begin{table}[H]
    \centering
    \scalebox{0.95}{%
    \begin{tabular}{|c|c|c|c|c|c|}\hline
        $\sA \otimes \Arf$ & $\sA$ & $Q_{\sA}$ & $(-1)^{\sF_{\Arf}}$ & $s_{\sF}$ & $\sF$\\ \hline\hline
        \cellcolor{Scolor}$\sS \otimes \mathsf{AP}$ & \multirow{4}{4em}{untwisted $u_{\sA}=+1$} & $ \cellcolor{Scolor}+1$ &  \cellcolor{Scolor}$ \cellcolor{Scolor}+1$ &  \cellcolor{Scolor}$-1$ &  \cellcolor{Scolor}bosonic, antiperiodic $(\sS, \mathsf{AP}_{\sF})$ \\ 
        \cline{1-1}\cline{3-6}$\sS \otimes \mathsf{P}$ & & $+1$ & $-1$ & $\times$ & $\times$ \\
        \cline{1-1}\cline{3-6}$\sT \otimes \mathsf{AP}$ & & $-1$ & $+1$ & $\times$ & $\times$ \\ 
        \cline{1-1}\cline{3-6}\cellcolor{Tcolor}$\sT \otimes \mathsf{P}$ & & \cellcolor{Tcolor}$-1$ & \cellcolor{Tcolor}$-1$ & \cellcolor{Tcolor}$+1$ & \cellcolor{Tcolor}fermionic, periodic $(\sT, \mathsf{P}_{\sF})$ \\ \hline\hline
        \cellcolor{Ucolor}$\sU \otimes \mathsf{AP}$ & \multirow{4}{4em}{twisted $u_{\sA}=-1$} &  \cellcolor{Ucolor}$+1$ &  \cellcolor{Ucolor}$+1$ &  \cellcolor{Ucolor}$+1$ &  \cellcolor{Ucolor}bosonic, periodic $(\sU, \mathsf{P}_{\sF})$ \\ 
        \cline{1-1}\cline{3-6}$\sU \otimes \mathsf{P}$ & & $+1$ & $-1$ & $\times$ & $\times$ \\
        \cline{1-1}\cline{3-6}$\sV \otimes \mathsf{AP}$ & & $-1$ & $+1$ & $\times$ & $\times$ \\ 
        \cline{1-1}\cline{3-6} \cellcolor{Vcolor}$\sV \otimes \mathsf{P}$ & & \cellcolor{Vcolor}$-1$ & \cellcolor{Vcolor}$-1$ & \cellcolor{Vcolor}$-1$ & \cellcolor{Vcolor}fermionic, antiperiodic $(\sV, \mathsf{AP}_{\sF})$ \\ \hline
    \end{tabular}
    }
    \caption{\label{tbl:stateF}States of the theories $\sA \otimes \Arf$ and $\sF$. Note that $s_{\sF} = -(-1)^{\sF_{\Arf}}u_{\sA}$.}
\end{table}
Fermionizing $\sD$ using $\Arf$ leads to a new fermionic theory:
\eqa{
    \sF' &:= (\sD \otimes \Arf)\orb\mathsf{diag}(\bZ_{2}^{(\sD)}\times \bZ_{2}^{\Arf}) ~.
}
Since $\sT$ and $\sU$ are exchanged in going from theory $\sA$ to theory $\sD$, the states of the theory $\sF'$ can be inferred directly from those of $\sF$, see Table \ref{tbl:stateFandFtilde}. If $\sU$ and $\sT$ are isomorphic, the theories $\sF$ and $\sF' = \sF \otimes \Arf$ coincide.
\begin{table}[H]
    \centering
    \scalebox{0.95}{%
    \begin{tabular}{|c|c|c|} \cline{2-3}
        \multicolumn{1}{c|}{$\sF$} & antiperiodic ($\sN\sS$) & periodic ($\sR$) \\ \hline
        bosonic & \cellcolor{Scolor}{$\sS$} & \cellcolor{Ucolor}{$\sU$} \\ \hline
        fermionic & \cellcolor{Vcolor}{$\sV$} & \cellcolor{Tcolor}{$\sT$} \\ \hline
    \end{tabular}
    \quad
    \begin{tabular}{|c|c|c|} \cline{2-3}
        \multicolumn{1}{c|}{$\sF'$} & antiperiodic ($\sN\sS$) & periodic ($\sR$) \\ \hline
        bosonic & \cellcolor{Scolor}{$\sS$} & \cellcolor{Tcolor}{$\sT$} \\ \hline
        fermionic & \cellcolor{Vcolor}{$\sV$} & \cellcolor{Ucolor}{$\sU$} \\ \hline
    \end{tabular}
    }
    \caption{\label{tbl:stateFandFtilde}States of theories $\sF$ and $\sF'$.}
\end{table}

Our remarks so far applied to 2d field theories. We now assume that $\sA$ is a 2d bosonic CFT with central charge $(c_L, c_R)$. The torus partition function contributions from the $\sS$, $\sT$, $\sU$, and $\sV$ sectors are
\eqa{
    \sZ_{\sS} &= \mathsf{tr}_{\mathsf{untwisted}}\left(\frac{1+\sfg}{2}\right)q^{L_0 - c_L/24}\ov{q}^{\ov{L}_0 - c_R/24} ~, \quad &\sZ_{\sT} &= \mathsf{tr}_{\mathsf{untwisted}}\left(\frac{1-\sfg}{2}\right)q^{L_0 - c_L/24}\ov{q}^{\ov{L}_0 - c_R/24} ~,\\
    \sZ_{\sU} &= \mathsf{tr}_{\mathsf{twisted}}\left(\frac{1+\sfg}{2}\right)q^{L_0 - c_L/24}\ov{q}^{\ov{L}_0 - c_R/24} ~, \quad &\sZ_{\sV} &= \mathsf{tr}_{\mathsf{twisted}}\left(\frac{1-\sfg}{2}\right)q^{L_0 - c_L/24}\ov{q}^{\ov{L}_0 - c_R/24} ~,
}
where $\sfg$ denotes the nontrivial generator of the $\IZ_2^{(\sA)}$ action on the Hilbert space. Note that $\sZ_{\sS}$, $\sZ_{\sT}$, $\sZ_{\sU}$ and $\sZ_{\sV}$ have \underline{all} non-negative $q, \ov{q}$-expansion coefficients. 

\noindent Under an $\mathcal{S}$ transformation ($\tau \mapsto -\frac{1}{\tau}$), 
\eqa{
    \sZ_{\sA} = \sZ_{\sS} + \sZ_{\sT} &\xlongrightarrow{\mathcal{S}} \sZ_{\sU} + \sZ_{\sV} ~, \qquad
    \sZ_{\sD} = \sZ_{\sS} + \sZ_{\sU} &\xlongrightarrow{\mathcal{S}} \sZ_{\sT} + \sZ_{\sV} ~.
}
From Table \ref{tbl:stateF}, we have, for the fermionic theory $\sF$,
\eqa{
    \sZ^{\sF}_{\sN\sS} &= \sZ_{\sS} + \sZ_{\sV} ~, \qquad
    \sZ^{\sF}_{\sR} &= \sZ_{\sU} + \sZ_{\sT} ~.
}
The torus partition functions of a fermionic theory are graded by the choice of spin structure $\sN\sS$ (antiperiodic) or $\sR$ (periodic) on each of the two homologically nontrivial $1$-cycles \cite{Alvarez-Gaume:1986rcs}:
\eqa{
 &\sZ_{\sN\sS,\sN\sS} &&=  \mathsf{tr}_{\sN\sS}\, q^{L_0 -\frac{c_L}{24}} \ov{q}^{\ov{L}_0 -\frac{c_R}{24}} ~, \label{eq:ferm00} \\
 &\sZ_{\sN\sS,\sR} &&=  \mathsf{tr}_{\sN\sS}\, (-1)^{F} q^{L_0 -\frac{c_L}{24}} \ov{q}^{\ov{L}_0 -\frac{c_R}{24}} ~, \label{eq:ferm01}\\
 &\sZ_{\sR,\sN\sS} &&=  \mathsf{tr}_{\sR}\, q^{L_0 -\frac{c_L}{24}} \ov{q}^{\ov{L}_0 -\frac{c_R}{24}} ~, \label{eq:ferm10}\\
 &\sZ_{\sR,\sR} &&=  \mathsf{tr}_{\sR}\, (-1)^{F} q^{L_0 -\frac{c_L}{24}} \ov{q}^{\ov{L}_0 -\frac{c_R}{24}} ~. \label{eq:ferm11}
}
By design, $\sZ_{\sN\sS,\sN\sS}$ and $\sZ_{\sR,\sN\sS}$ have \underline{all} nonnegative coefficients, whereas the other two do not. Also, $\sZ_{\sR,\sN\sS}$ and $\sZ_{\sN\sS,\sR}$ are related by an $\mathcal{S}$ transformation. Therefore, $\sZ^{\sF}_{\sN\sS}$ can be identified with $\sZ^{\sF}_{\sN\sS,\sN\sS}$ and  $\sZ^{\sF}_{\sR}$ with $\sZ^{\sF}_{\sR,\sN\sS}$. To summarize, the partition functions of $\sF$ and $\sF'$ can be reconstructed from those of $\sA$ as follows:\\
\begin{minipage}{0.5\textwidth}
\eqa{
& \sZ^{\sF}_{\sN\sS,\sN\sS} &&= \sZ_\sS + \sZ_\sV ~, \label{eq:fermNS-NS}\\
& \sZ^{\sF}_{\sN\sS,\sR} &&= \sZ_\sS - \sZ_\sV ~, \label{eq:fermNS-R}\\
& \sZ^{\sF}_{\sR,\sN\sS} &&= \sZ_\sU + \sZ_\sT ~, \label{eq:fermR-NS}\\
& \sZ^{\sF}_{\sR,\sR} &&= \sZ_\sU - \sZ_\sT ~. \label{eq:fermR-R}
}
\end{minipage}%
\begin{minipage}{0.5\textwidth}
\eqa{
& \sZ^{\sF'}_{\sN\sS,\sN\sS} &&= \sZ_\sS + \sZ_\sV ~, \label{eq:Fprime-fermNS-NS}\\
& \sZ^{\sF'}_{\sN\sS,\sR} &&= \sZ_\sS - \sZ_\sV ~, \label{eq:Fprime-fermNS-R}\\
& \sZ^{\sF'}_{\sR,\sN\sS} &&= \sZ_\sU + \sZ_\sT ~, \label{eq:Fprime-fermR-NS}\\
& \sZ^{\sF'}_{\sR,\sR} &&= -\sZ_\sU + \sZ_\sT ~. \label{eq:Fprime-fermR-R}
}
\end{minipage}\\
\\
The $\Arf$ theory is 2d spin CFT with central charge $(c_L, c_R) = (0,0)$, and its torus partition function is
\eqa{
  \sZ_{p,q}^{\Arf} &= (-1)^{\mathsf{Arf}[\IT^2,(p,q)]} = \left\{
  \begin{array}{ll}
    +1 ~, & (p, q) = (\sN\sS,\sN\sS), (\sN\sS,\sR), (\sR,\sN\sS) ~,\\
    -1 ~, & (p, q) = (\sR,\sR) ~,
  \end{array}
  \right. \label{eq:Z-Arf-torus}
}
where the $\mathrm{Arf}$ invariant $\mathrm{Arf}[\IT^2,(p,q)]$ was defined in footnote \ref{foot:arfinvt}.

In \cite{BoyleSmith:2023xkd}, the symbols $\chi_{\sN\sS}^{\text{even}}$ and $\chi_{\sN\sS}^{\text{odd}}$ are used to denote $\sN\sS$ sector characters of fermionic CFTs corresponding with $(-1)^{F}$ even and odd respectively. Here is a translation between their notation and ours:\footnote{The theories in \cite{BoyleSmith:2023xkd} are chiral, so the right-moving sector is empty ($c_{R} = 0$) and their $\chi$'s denote chiral characters.}
\eqa{
    \chi_{\sN\sS}^{\text{even}} &= \frac{1}{2}\big( \sZ^{\sF}_{\sN\sS,\sN\sS} +  \sZ^{\sF}_{\sN\sS,\sR}) = ``\sZ_{\sS}" = \mathsf{tr}_{\sN\sS}\,\left(\frac{1 + (-1)^{F}}{2}\right) q^{L_0 -\frac{c_L}{24}} \ov{q}^{\ov{L}_0 -\frac{c_R}{24}}  ~,\\
    \chi_{\sN\sS}^{\text{odd}} &= \frac{1}{2}\big( \sZ^{\sF}_{\sN\sS,\sN\sS} -  \sZ^{\sF}_{\sN\sS,\sR}) = ``\sZ_{\sV}" = \mathsf{tr}_{\sN\sS}\,\left(\frac{1 - (-1)^{F}}{2}\right) q^{L_0 -\frac{c_L}{24}} \ov{q}^{\ov{L}_0 -\frac{c_R}{24}}  ~.
}
Clearly, under an $\cS$ transformation,
\eqa{
    \chi_{\sN\sS}^{\text{even}} -  \chi_{\sN\sS}^{\text{odd}} &= \sZ_{\sN\sS,\sR}^{\sF} \xlongrightarrow{\cS} \sZ^{\sF}_{\sR,\sN\sS} ~,
}
gives the partition function on a torus in the $(\sR,\sN\sS)$ sector.\footnote{In \cite{BoyleSmith:2023xkd}, this is referred to as the torus partition function with the spatial circle in the $\sR$ sector and the temporal circle in the $\sN\sS$ (antiperiodic) sector. In our notation, the first subscript corresponds to the spin structure around the ``spatial'' circle, and the second corresponds to that around the ``temporal'' circle, denoted there by $\mathcal{Z}_{\text{space}}^{\text{time}}$.}

If $(c_L - c_R)$ is strictly integral, $(-1)^{F}$ is well-defined on the $\sR$ sector, and we have,
\eqa{
    \chi_{\sR}^{\text{even}} &= \frac{1}{2}\big( \sZ^{\sF}_{\sR,\sN\sS} +  \sZ^{\sF}_{\sR,\sR}) = ``\sZ_{\sU}" =  \mathsf{tr}_{\sR}\,\left(\frac{1 + (-1)^{F}}{2}\right) q^{L_0 -\frac{c_L}{24}} \ov{q}^{\ov{L}_0 -\frac{c_R}{24}} ~,\\
    \chi_{\sR}^{\text{odd}} &= \frac{1}{2}\big( \sZ^{\sF}_{\sR,\sN\sS} -  \sZ^{\sF}_{\sR,\sR}) = ``\sZ_{\sT}" =  \mathsf{tr}_{\sR}\,\left(\frac{1 - (-1)^{F}}{2}\right) q^{L_0 -\frac{c_L}{24}} \ov{q}^{\ov{L}_0 -\frac{c_R}{24}} ~.
}
There is a variant of this obtained by tensoring with the $\mathsf{Arf}$ theory, which exchanges $\chi_{\sR}^{\text{even}}$ and $\chi_{\sR}^{\text{odd}}$.

If $(c_L - c_R)$ is strictly half-integral, the resulting partition function computed by $\cS(\chi_{\sN\sS}^{\text{even}} -  \chi_{\sN\sS}^{\text{odd}}) = \sZ^{\sF}_{\sR,\sN\sS}$ has non-negative $q$-expansion coefficients but the coefficients are $\sqrt{2}$ times non-negative integers. It is conventional to write this as $\sqrt{2}\chi_{\sR}$, where $\chi_{\sR}$ has non-negative integer $q$-expansion coefficients. In this case, the fermionic parity $(-1)^{F}$ is not well-defined on the states counted by $\chi_{\sR}$. While this is true in general whenever $(c_L - c_R)$ is strictly half-integral, one extreme example of it is the theory of the single Majorana-Weyl fermion, see \cite{Witten:2023snr,Tachikawa:2023nne,Freed:2024apc} -- we will encounter it in Section \ref{sec:tachyon}. In these cases, $(-1)^{F}$ is not well-defined in the $\sR$ sector and consequently, $\chi_{\sR}^{\text{even}}$ and $\chi_{\sR}^{\text{odd}}$ are not well-defined.


\section{Theories from Fermionization}\label{sec:fermionizedtheories}
In this section, we will assemble the partition functions of various building block theories, leading up to the partition functions of theories $\sT_1$ and $\sT_2$. We will extensively use the methods of Section \ref{sec:modernz2orbifoldintro}.

\subsection{Compact Boson}
The (free) compact boson CFT $X$ on the 2-torus is a map $X: \IT^{2} \to S^{1}_{R}$ (denoted by $\cbos_{R}$), with action,
\eqa{
  S(\cbos_{R}) &= \frac{1}{4\pi}\int_{\IT^2}d^{2}z \partial X \ov{\partial} X ~.
}
It has central charge $(c_L, c_R) = (1,1)$.  The partition function on $\IT^{2}$ with modular parameter $\tau$ is
\eqa{
 \sZ(\cbos_{R}) := \frac{1}{|\eta(\tau)|^2}\sum_{m,n \in \bZ} q^{\frac{1}{4}\left(\frac{m}{R} + n R\right)^2}\ov{q}^{\frac{1}{4}\left(\frac{m}{R} - n R\right)^2} ~, \label{eq:pf-cboson}
} 
where $q = e^{2\pi \sfi\tau}$, $m$ and $n$ are the momentum and winding modenumbers, and $\eta(\tau)$ is the Dedekind function (see Appendix \ref{app:ThetaFunc}). In our conventions, $\alpha' = 1$. The partition function satisfies $\sZ(\cbos_{R}) = \sZ(\cbos_{1/R})$ -- this is the familiar T-Duality of the compact boson.

The CFT $\cbos_{R}$ has a non-anomalous $\IZ_2$ symmetry, under which $X$ shifts by half a period: $X \mapsto X + \pi R$. We denote it by $\IZ_2^{\hp}$. This acts on the Hilbert space state $|m,n\rangle$ as $|m, n\rangle \mapsto (-1)^{m}|m, n\rangle$. The $\IZ_2^\hp$-twisted partition function is written as $\sZ_{\sfa,\sfb}$, where $\sfa$ and $\sfb$ denote the $\IZ_2$-holonomies around the two homologically nontrivial 1-cycles of $\IT^{2}$. Specifically, the $\IZ_2^\hp$-twisted partition function of $\cbos_{R}$ is
\eqa{
  & \sZ_{\mathsf{a,b}}(\cbos_{R}; \IZ_{2}^{\hp}) &&= \frac{1}{|\eta(\tau)|^2}\sum_{\substack{m \in \bZ \\ n \in \bZ + \frac{1}{2}\mathsf{a}}}(-1)^{m \mathsf{b}}q^{\frac{1}{4}\left(\frac{m}{R} + nR\right)^2}\ov{q}^{\frac{1}{4}\left(\frac{m}{R} - nR\right)^2} ~. \label{eq:z-cbosR-ab}
}
In the $\IZ_2^\hp$-twisted sector, the winding number is valued in $\IZ + \frac{1}{2}$. Taking $\cbos_{2R}$ as `theory $\sA$' in the language of Section \ref{sec:modernz2orbifoldintro}, and using \eqref{eq:z-cbosR-ab} the $\sS$, $\sT$, $\sU$, $\sV$ decompositions are
\eqa{
& \sZ_{\sS}(\cbos_{2R}; \IZ_{2}^{\hp}) &&= \frac{1}{2|\eta(\tau)|^2}\sum_{m,n \in \bZ}\left(1 + (-1)^{m}\right)q^{\frac{1}{4}\left(\frac{m}{2R} + 2nR\right)^2}\ov{q}^{\frac{1}{4}\left(\frac{m}{2R} - 2nR\right)^2} ~, \label{eq:cbos-2R-S}\\
& \sZ_{\sT}(\cbos_{2R}; \IZ_{2}^{\hp}) &&= \frac{1}{2|\eta(\tau)|^2}\sum_{m,n \in \bZ}\left(1 - (-1)^{m}\right)q^{\frac{1}{4}\left(\frac{m}{2R} + 2nR\right)^2}\ov{q}^{\frac{1}{4}\left(\frac{m}{2R} - 2nR\right)^2} ~, \label{eq:cbos-2R-T}\\
& \sZ_{\sU}(\cbos_{2R}; \IZ_{2}^{\hp}) &&= \frac{1}{2|\eta(\tau)|^2}\sum_{m,n \in \bZ}\left(1 + (-1)^{m}\right)q^{\frac{1}{4}\left(\frac{m}{2R} + 2\left(n+\frac{1}{2}\right)R\right)^2}\ov{q}^{\frac{1}{4}\left(\frac{m}{2R} - 2\left(n+\frac{1}{2}\right)R\right)^2} ~, \label{eq:cbos-2R-U}\\
& \sZ_{\sV}(\cbos_{2R}; \IZ_{2}^{\hp}) &&= \frac{1}{2|\eta(\tau)|^2}\sum_{m,n \in \bZ}\left(1 - (-1)^{m}\right)q^{\frac{1}{4}\left(\frac{m}{2R} + 2\left(n+\frac{1}{2}\right)R\right)^2}\ov{q}^{\frac{1}{4}\left(\frac{m}{2R} - 2\left(n+\frac{1}{2}\right)R\right)^2} ~. \label{eq:cbos-2R-V}
}
The orbifold $\sD(\cbos_{2R}) := \cbos_{2R}\orb\IZ_{2}^{\hp} =\cbos_{R}$ is a compact boson at half the radius, with partition function
\eqa{
& \sZ\big(\sD(\cbos_{2R})\big) &&= \sZ_{\sS}(\cbos_{2R}; \IZ_{2}^{\hp}) + \sZ_{\sU}(\cbos_{2R}; \IZ_{2}^{\hp}) = \sZ(\cbos_{R}) ~.
}
The fermionization,
\eqa{
  \sF(\cbos_{2R}; \IZ_{2}^{\hp} ) := (\cbos_{2R} \otimes \Arf)\orb\mathsf{diag}(\IZ_{2}^{\hp} \times \IZ_{2}^{\Arf}) ~,
 }
has partition functions given by \eqref{eq:fermNS-NS}--\eqref{eq:fermR-R}:
\eqa{
  &\sZ_{\sN\sS,\sN\sS}\big(\sF(\cbos_{2R}; \IZ_{2}^{\hp})\big) &= \sZ_{\sS}(\cbos_{2R}; \IZ_{2}^{\hp}) + \sZ_{\sV}(\cbos_{2R}; \IZ_{2}^{\hp}) ~, \label{eq:cbos-ferm-NSNS}\\
  &\sZ_{\sN\sS,\sR}\big(\sF(\cbos_{2R}; \IZ_{2}^{\hp}) \big) &= \sZ_{\sS}(\cbos_{2R}; \IZ_{2}^{\hp}) - \sZ_{\sV}(\cbos_{2R}; \IZ_{2}^{\hp}) ~,\label{eq:cbos-ferm-NSR}\\
  &\sZ_{\sR,\sN\sS}\big(\sF(\cbos_{2R}; \IZ_{2}^{\hp}) \big) &= \sZ_{\sU}(\cbos_{2R}; \IZ_{2}^{\hp}) + \sZ_{\sT}(\cbos_{2R}; \IZ_{2}^{\hp}) ~,\label{eq:cbos-ferm-RNS}\\
  &\sZ_{\sR,\sR}\big(\sF(\cbos_{2R}; \IZ_{2}^{\hp}) \big) &= \sZ_{\sU}(\cbos_{2R}; \IZ_{2}^{\hp}) - \sZ_{\sT}(\cbos_{2R}; \IZ_{2}^{\hp}) ~. \label{eq:cbos-ferm-RR}
}
The fermionization of the compact boson by the $\Arf$ theory has also been discussed in \cite{Okuda:2020fyl}. Note that \eqref{eq:cbos-ferm-NSNS}--\eqref{eq:cbos-ferm-RR} can be written as:
\eqa{
  &\sZ_{p,q}\big(\sF(\cbos_{2R}; \IZ_{2}^{\hp})\big) 
  &&= \frac{1}{2|\eta(\tau)|^2}\sum_{\substack{s_1, s_2 \\ \in \{\sN\sS, \sR\}} }(-1)^{(s_1+p)(s_2+q)}\sum_{\substack{m\in \IZ \\ n \in \IZ + \frac{1}{2}s_1}} (-1)^{m s_2}q^{\frac{1}{4}\left(\frac{m}{2R} + 2nR\right)^2}\ov{q}^{\frac{1}{4}\left(\frac{m}{2R} - 2nR\right)^2} ~,
  %
} 
where $p, q \in \{\sN\sS, \sR\}$ and we identify\footnote{Note that $s_1 + p$ and $s_2 + q$ are defined by the relations $\sN\sS + \sN\sS = \sN\sS$ = 0, $\sN\sS + \sR = \sR + \sN\sS = \sR = 1$, and $\sR + \sR = \sN\sS = 0$.} $\{\sN\sS, \sR\} = \{0, 1\}$. 

\subsection{$(E_8)_{1} \times (E_8)_{1}$ and $(E_8)_2 \times \lambda$}
The $(E_8)_1 \times (E_8)_1$ current algebra as a holomorphic (left-moving) CFT has central charge $(c_L, c_R) = (16,0)$. It admits a non-anomalous $\IZ_2$ symmetry that exchanges the two $E_8$ factors, denoted by $\IZ_{2}^{\sigma}$. Using the coset relation,
\eqa{
  \frac{(E_8)_1 \times (E_8)_1}{(E_8)_2} = \Vir_{c=1/2} ~,
}
where $\mathsf{Vir}_{c=1/2}$ denotes the holomorphic theory of a Majorana-Weyl fermion (sometimes denoted by $\lambda$), each state in a representation of $(E_8)_1 \times (E_8)_1$ can be written as a product of a state in a representation of $(E_8)_2$ with a state in a representation of $\mathsf{Vir}_{c=1/2}$  \cite[Ch. 18]{DiFrancesco:1997nk}. In particular, the $\IZ_{2}^{\sigma}$-twisted partition functions of $(E_8)_1 \times (E_8)_1$ can be decomposed in terms of the characters of $(E_8)_{2}$,
\eqa{
  \chi_{\bm{1}}^{E_8} ~, \quad \chi_{\bm{248}}^{E_8} ~, \quad \chi_{\bm{3875}}^{E_8} ~,
}
and the characters of $\Vir_{c=1/2}$,
\eqa{
  \chi_{0}^{\Vir} ~, \quad \chi_{1/2}^{\Vir}, ~\quad \chi_{1/16}^{\Vir} ~.
}
 (See Appendix \ref{app:compchar} and \cite{Forgacs:1988iw} for details.) The $\IZ_{2}^{\sigma}$-twisted partition functions of $(E_8)_1 \times (E_8)_1$ are
\eqa{
  &\sZ_{\mathsf{0,0}}\big( (E_8)_1 \times (E_8)_1; \IZ_{2}^{\sigma} \big) &&= \chi_{0}^{\Vir}\chi_{\bm{1}}^{E_8} + \chi_{1/2}^{\Vir}\chi_{\bm{3875}}^{E_8} + \chi_{1/16}^{\Vir}\chi_{\bm{248}}^{E_8} ~,\\
  &\sZ_{\mathsf{0,1}}\big( (E_8)_1 \times (E_8)_1; \IZ_{2}^{\sigma} \big) &&= \chi_{0}^{\Vir}\chi_{\bm{1}}^{E_8} + \chi_{1/2}^{\Vir}\chi_{\bm{3875}}^{E_8} - \chi_{1/16}^{\Vir}\chi_{\bm{248}}^{E_8} ~,\\
  &\sZ_{\mathsf{1,0}}\big( (E_8)_1 \times (E_8)_1; \IZ_{2}^{\sigma} \big) &&= \chi_{1/2}^{\Vir}\chi_{\bm{1}}^{E_8} + \chi_{0}^{\Vir}\chi_{\bm{3875}}^{E_8} + \chi_{1/16}^{\Vir}\chi_{\bm{248}}^{E_8} ~,\\
  &\sZ_{\mathsf{1,1}}\big( (E_8)_1 \times (E_8)_1; \IZ_{2}^{\sigma} \big) &&= -\chi_{1/2}^{\Vir}\chi_{\bm{1}}^{E_8} - \chi_{0}^{\Vir}\chi_{\bm{3875}}^{E_8} + \chi_{1/16}^{\Vir}\chi_{\bm{248}}^{E_8} ~.
}
So the $\sS$, $\sT$, $\sU$, $\sV$ decompositions are
\eqa{
  &\sZ_{\sS}\big( (E_8)_1 \times (E_8)_1; \IZ_{2}^{\sigma} \big) &&= \chi_{0}^{\Vir}\chi_{\bm{1}}^{E_8} + \chi_{1/2}^{\Vir}\chi_{\bm{3875}}^{E_8} ~, \label{eq:E8byE8-S}\\
  &\sZ_{\sT}\big( (E_8)_1 \times (E_8)_1; \IZ_{2}^{\sigma} \big) &&= \chi_{1/16}^{\Vir}\chi_{\bm{248}}^{E_8} ~, \label{eq:E8byE8-T}\\
  &\sZ_{\sU}\big( (E_8)_1 \times (E_8)_1; \IZ_{2}^{\sigma} \big) &&= \chi_{1/16}^{\Vir}\chi_{\bm{248}}^{E_8} ~, \label{eq:E8byE8-U}\\
  &\sZ_{\sV}\big( (E_8)_1 \times (E_8)_1; \IZ_{2}^{\sigma} \big) &&= \chi_{1/2}^{\Vir}\chi_{\bm{1}}^{E_8} + \chi_{0}^{\Vir}\chi_{\bm{3875}}^{E_8} ~. \label{eq:E8byE8-V}
}
The fermionization of this theory is referred to as $(E_8)_{2} \times \mathsf{Ising}$ or $(E_8)_{2} \times \lambda$,
\eqa{
  \sF\big( (E_8)_1 \times (E_8)_1; \IZ_{2}^{\sigma} \big) :=  (E_8)_{2} \times \lambda = \big( (E_8)_1 \times (E_8)_1 \otimes \Arf\big)\orb \mathsf{diag}( \IZ_{2}^{\sigma} \times \IZ_{2}^{\Arf}) ~, \label{eq:e8-level2-psi}
}
and has the following partition functions:
\eqa{
  &\sZ_{\sN\sS,\sN\sS}\big((E_8)_{2} \times \lambda \big) &&= \big(\chi_{0}^{\Vir} + \chi_{1/2}^{\Vir}\big)\big(\chi_{\bm{1}}^{E_8} + \chi_{\bm{3875}}^{E_8}\big) ~, \label{eq:e8-level2-psi-NSNS}\\
  &\sZ_{\sN\sS,\sR}\big((E_8)_{2} \times \lambda \big) &&= \big(\chi_{0}^{\Vir} - \chi_{1/2}^{\Vir}\big)\big(\chi_{\bm{1}}^{E_8} - \chi_{\bm{3875}}^{E_8}\big) ~, \label{eq:e8-level2-psi-NSR}\\
   &\sZ_{\sR,\sN\sS}\big((E_8)_{2} \times \lambda \big) &&= \sqrt{2}\chi_{1/16}^{\Vir} \cdot \sqrt{2}\chi_{\bm{248}}^{E_8} ~, \label{eq:e8-level2-psi-RNS}\\
   &\sZ_{\sR,\sR}\big((E_8)_{2} \times \lambda \big) &&= 0 ~. \label{eq:e8-level2-psi-RR}
}
The fermionic theory $(E_8)_2 \times \lambda$ has a $\IZ_2$ symmetry, generated by the fermion number operator $(-1)^{F}$. Note that the $\sT$ and $\sU$ sectors of $(E_8)_1 \times (E_8)_1$ are identical, so the fermionization $\sF' = \sF \otimes \Arf$ coincides with $\sF$.

\subsection{Intermezzo: Gauging Fermion Parity}\label{sec:fermparity}
Defining a fermionic theory $\sF$ (with a $\IZ_2$ fermion parity symmetry generated by the fermion number $(-1)^{F}$) on a Riemann surface $\Sigma$ requires a choice of spin structure $\rho$. We denote its partition function by $\sZ_{\sF}[\Sigma,\rho]$. If we turn on a gauge field $\bm{\sfa}$ for $(-1)^{F}$, it acts by shifting the spin structure: $\rho \mapsto \rho + \bm{\sfa}$. Equivariantizing the theory with a fixed spin structure $\rho$ with respect to $(-1)^{F}$ amounts to treating $\bm{\sfa}$ as a fluctuating field that is now summed over, yielding a theory $\widetilde{\sF}$ with partition function
\eqa{
  \sZ_{\widetilde{\sF}}[\Sigma,\rho,A] &= \frac{1}{\sqrt{\left|H^1(\Sigma,\IZ_2)\right|}}\sum_{\bm{\sfa}}(-1)^{\int_{\Sigma}\bm{\sfa}\,\cup A}\sZ_{\sF}[\rho + \bm{\sfa}] ~, \label{eq:zftilde}
}
where we have turned on a background field $A$ for the $\IZ_2$ symmetry of $\widetilde{\sF}$, and the phase factor has an expression in terms of Arf invariants, details of which can be found in \cite{Karch:2019lnn}.\footnote{Summing over the gauge field $\sfa$ naively seems to be equivalent to summing over spin structures, but a careful analysis \cite{BoyleSmith:2023xkd,BoyleSmith:2024qgx} reveals that spin structure summation and gauging fermion parity are in general different operations, and coincide only when the gravitational anomaly of the theory is $0 \mathsf{\,\,mod\,\,}16$: in such cases the theory $\widetilde{F}$ with partition function \eqref{eq:zftilde} is a bosonic theory (this is true bosonization). Gauging fermion parity is possible even when the anomaly is $8 \mathsf{\,\,mod\,\,}16$, in which case the result is another fermionic theory (refermionization).}

Gauging fermion parity can be implemented by turning on holonomies for $\IZ_2$ gauge fields (around the $A$- and $B$- cycles of $\IT^2$), represented by a pair of integers $\sfa, \sfb \in \{0, 1\}$, while retaining the spin structure grading on $\IT^2$ given by $(p, q)$ where $p, q \in \{\sN\sS, \sR\}$. This leads to a ``refined'' partition function labeled by four indices, $\bm{\sZ}^{\sF}_{p, \sfa; q, \sfb}$. The background gauge field $\bm{\sfa}$ for $(-1)^{F}$ shifts the spin structure, leading to
\eqa{
  \bm{\sZ}^{\sF}_{p, \sfa; q, \sfb} &:= \sZ^{\sF}_{p+\sfa,q+\sfb} ~, \label{eq:zpec-concept}
}
where the RHS is completely specified by the unrefined fermionic partition function and the $\mathsf{mod\,}2$ rules that define $p + \sfa$.\footnote{$\sN\sS + \mathsf{0} = \sN\sS$, $\sN\sS + \mathsf{1} = \sR$, $\sR + \mathsf{0} = \sR$, and $\sR + \mathsf{1} = \sN\sS$.} One can consider the same operation on $\sF' = \sF \otimes \Arf$ instead of $\sF$. In this case, 
\eqa{
& \bm{\sZ}^{\sF'}_{p, \sfa; q, \sfb}  &&:= \sZ_{p+\sfa,q+\sfb}^{\sF}\sZ^{\Arf}_{p,q} ~,  \label{eq:curious_z2_gen}
}
where $\sZ_{p,q}^{\Arf}$ is given by \eqref{eq:Z-Arf-torus}. Given a bosonic theory $\cT_{\mathsf{bos}}$ with a non-anomalous $\IZ_2^{(\mathsf{bos})}$-symmetry, one can construct a tensor product $\cT_{\mathsf{bos}} \otimes \sF$ and gauge it by a $\IZ_2$-action that acts as $\IZ_{2}^{(\mathsf{bos})}$ on $\cT_{\mathsf{bos}}$ and as $(-1)^{F}$ on $\sF$. The $\IT^{2}$-partition function of the orbifold theory,
\eqa{
  \big(\cT_{\mathsf{bos}}\otimes \sF\big)\orb\mathsf{diag}\big(\IZ_2^{(\mathsf{bos})} \times (-1)^{F}\big) ~, \label{eq:gaugingbosF}
}
is given by\footnote{For $\Sigma = \IT^{2}$, $\left|H^1(\Sigma,\IZ_2)\right| = 4$, so one gets a familiar factor of $1/2$.}
\eqa{
  &\sZ^{}_{p,q} &&= \frac{1}{2}\sum_{\sfa,\sfb\in\{0,1\}}\sZ^{\cT_{\mathsf{bos}}}_{\sfa,\sfb}\bm{\sZ}^{\sF}_{p,\sfa;q,\sfb} = \frac{1}{2}\sum_{\sfa,\sfb\in\{0,1\}}\sZ^{\cT_{\mathsf{bos}}}_{\sfa,\sfb}\sZ^{\sF}_{p+\sfa,q+\sfb} ~. \label{eq:zpec}
}
One can construct another orbifold by replacing $\sF$ with $\sF'$ in \eqref{eq:gaugingbosF}. 

\subsection{Tensoring $(E_8)_1 \times (E_8)_1$ and $(E_8)_2 \times \lambda$ with a Compact Boson}\label{sec:tensoringe8}
Following Section \ref{sec:modernz2orbifoldintro}, we take as `theory $\sA$' the tensor product,
\eqa{
  \cA_1 := \cbos_{2R} \otimes (E_8)_1 \times (E_8)_1 ~. \label{eq:Theory-CA1}
}
This theory admits a $\IZ_2^\hp \times \IZ_2^\sigma$ symmetry. The twisted partition function with holonomies $(\mathsf{a,b})$ is
\eqa{
& \sZ_{\mathsf{a,b}}\big( \cbos_{2R} \otimes (E_8)_1 \times (E_8)_1 ;\IZ_2^\hp \times \IZ_2^\sigma\big) &&:= \sZ_{\mathsf{a,b}}(\cbos_{2R}; \IZ_{2}^{\hp}) \times \sZ_{\mathsf{a,b}}((E_8)_1 \times (E_8)_1; \IZ_2^\sigma)
~.
}
For brevity, we define
\eqa{
  \sZ^{\mathsf{cb}}_{\mathsf{a,b}} := \sZ_{\mathsf{a,b}}(\cbos_{2R}; \IZ_{2}^{\hp}) ~.
}
In terms of $\sS$, $\sT$, $\sU$, $\sV$ decompositions, we have,
\begin{align}
\hspace{-0.15in} \sZ_{\sS}(\cA_1) 
&= \frac{1}{2}\left[\sZ_{\mathsf{0,0}}^{\mathsf{cb}} \left( \chi_{0}^{\Vir}\chi_{\bm{1}}^{E_8} + \chi_{1/2}^{\Vir}\chi_{\bm{3875}}^{E_8} + \chi_{1/16}^{\Vir}\chi_{\bm{248}}^{E_8} \right) + \sZ_{\mathsf{0,1}}^{\mathsf{cb}}\left( \chi_{0}^{\Vir}\chi_{\bm{1}}^{E_8} + \chi_{1/2}^{\Vir}\chi_{\bm{3875}}^{E_8} - \chi_{1/16}^{\Vir}\chi_{\bm{248}}^{E_8}  \right)\right] ~,\\
\hspace{-0.15in} \sZ_{\sT}(\cA_1) 
&= \frac{1}{2}\left[\sZ_{\mathsf{0,0}}^{\mathsf{cb}} \left( \chi_{0}^{\Vir}\chi_{\bm{1}}^{E_8} + \chi_{1/2}^{\Vir}\chi_{\bm{3875}}^{E_8} + \chi_{1/16}^{\Vir}\chi_{\bm{248}}^{E_8} \right) - \sZ_{\mathsf{0,1}}^{\mathsf{cb}}\left( \chi_{0}^{\Vir}\chi_{\bm{1}}^{E_8} + \chi_{1/2}^{\Vir}\chi_{\bm{3875}}^{E_8} - \chi_{1/16}^{\Vir}\chi_{\bm{248}}^{E_8}  \right)\right] ~,\\
\hspace{-0.15in} \sZ_{\sU}(\cA_1) 
 &= \frac{1}{2}\left[ \sZ_{\mathsf{1,0}}^{\mathsf{cb}}\left(\chi_{1/2}^{\Vir}\chi_{\bm{1}}^{E_8} + \chi_{0}^{\Vir}\chi_{\bm{3875}}^{E_8} + \chi_{1/16}^{\Vir}\chi_{\bm{248}}^{E_8}\right) + \sZ_{\mathsf{1,1}}^{\mathsf{cb}}\left(-\chi_{1/2}^{\Vir}\chi_{\bm{1}}^{E_8} - \chi_{0}^{\Vir}\chi_{\bm{3875}}^{E_8} + \chi_{1/16}^{\Vir}\chi_{\bm{248}}^{E_8}\right) \right]~,\\
\hspace{-0.15in} \sZ_{\sV}(\cA_1) 
&= \frac{1}{2}\left[ \sZ_{\mathsf{1,0}}^{\mathsf{cb}}\left(\chi_{1/2}^{\Vir}\chi_{\bm{1}}^{E_8} + \chi_{0}^{\Vir}\chi_{\bm{3875}}^{E_8} + \chi_{1/16}^{\Vir}\chi_{\bm{248}}^{E_8}\right) - \sZ_{\mathsf{1,1}}^{\mathsf{cb}}\left(-\chi_{1/2}^{\Vir}\chi_{\bm{1}}^{E_8} - \chi_{0}^{\Vir}\chi_{\bm{3875}}^{E_8} + \chi_{1/16}^{\Vir}\chi_{\bm{248}}^{E_8}\right) \right]~.
\end{align}
The fermionization of theory $\cA_1$,
\eqa{
  \sT_1 &:= \big( \cbos_{2R} \otimes (E_8)_1 \times (E_8)_1 \otimes \Arf\big)\orb\mathsf{diag}(\IZ_2^\hp \times \IZ_2^\sigma \times \IZ_2^\Arf) ~, \label{eq:Theory-CF1}
}
has the following partition functions:
\begin{align}
  \hspace{-0.15in}\sZ_{\sN\sS,\sN\sS}^{\sT_1} &= \frac{1}{2}\left[\sZ_{\mathsf{0,0}}^{\mathsf{cb}} \left( \chi_{0}^{\Vir}\chi_{\bm{1}}^{E_8} + \chi_{1/2}^{\Vir}\chi_{\bm{3875}}^{E_8} + \chi_{1/16}^{\Vir}\chi_{\bm{248}}^{E_8} \right) + \sZ_{\mathsf{0,1}}^{\mathsf{cb}}\left( \chi_{0}^{\Vir}\chi_{\bm{1}}^{E_8} + \chi_{1/2}^{\Vir}\chi_{\bm{3875}}^{E_8} - \chi_{1/16}^{\Vir}\chi_{\bm{248}}^{E_8}  \right)\right. \nonumber\\
 \hspace{-0.15in}&\qquad + \left. \sZ_{\mathsf{1,0}}^{\mathsf{cb}}\left(\chi_{1/2}^{\Vir}\chi_{\bm{1}}^{E_8} + \chi_{0}^{\Vir}\chi_{\bm{3875}}^{E_8} + \chi_{1/16}^{\Vir}\chi_{\bm{248}}^{E_8}\right) - \sZ_{\mathsf{1,1}}^{\mathsf{cb}}\left(-\chi_{1/2}^{\Vir}\chi_{\bm{1}}^{E_8} - \chi_{0}^{\Vir}\chi_{\bm{3875}}^{E_8} + \chi_{1/16}^{\Vir}\chi_{\bm{248}}^{E_8}\right) \right] ~, \label{eq:Z-F1-NSNS}\\
\hspace{-0.15in}\sZ_{\sN\sS,\sR}^{\sT_1} &= \frac{1}{2}\left[\sZ_{\mathsf{0,0}}^{\mathsf{cb}} \left( \chi_{0}^{\Vir}\chi_{\bm{1}}^{E_8} + \chi_{1/2}^{\Vir}\chi_{\bm{3875}}^{E_8} + \chi_{1/16}^{\Vir}\chi_{\bm{248}}^{E_8} \right) + \sZ_{\mathsf{0,1}}^{\mathsf{cb}}\left( \chi_{0}^{\Vir}\chi_{\bm{1}}^{E_8} + \chi_{1/2}^{\Vir}\chi_{\bm{3875}}^{E_8} - \chi_{1/16}^{\Vir}\chi_{\bm{248}}^{E_8}  \right)\right. \nonumber\\
 \hspace{-0.15in}&\qquad - \left. \sZ_{\mathsf{1,0}}^{\mathsf{cb}}\left(\chi_{1/2}^{\Vir}\chi_{\bm{1}}^{E_8} + \chi_{0}^{\Vir}\chi_{\bm{3875}}^{E_8} + \chi_{1/16}^{\Vir}\chi_{\bm{248}}^{E_8}\right) + \sZ_{\mathsf{1,1}}^{\mathsf{cb}}\left(-\chi_{1/2}^{\Vir}\chi_{\bm{1}}^{E_8} - \chi_{0}^{\Vir}\chi_{\bm{3875}}^{E_8} + \chi_{1/16}^{\Vir}\chi_{\bm{248}}^{E_8}\right) \right] ~, \label{eq:Z-F1-NSR}\\
\hspace{-0.15in}\sZ_{\sR,\sN\sS}^{\sT_1} &= \frac{1}{2}\left[ \sZ_{\mathsf{1,0}}^{\mathsf{cb}}\left(\chi_{1/2}^{\Vir}\chi_{\bm{1}}^{E_8} + \chi_{0}^{\Vir}\chi_{\bm{3875}}^{E_8} + \chi_{1/16}^{\Vir}\chi_{\bm{248}}^{E_8}\right) + \sZ_{\mathsf{1,1}}^{\mathsf{cb}}\left(-\chi_{1/2}^{\Vir}\chi_{\bm{1}}^{E_8} - \chi_{0}^{\Vir}\chi_{\bm{3875}}^{E_8} + \chi_{1/16}^{\Vir}\chi_{\bm{248}}^{E_8}\right) \right. \nonumber\\
 \hspace{-0.15in}&\qquad + \left. \sZ_{\mathsf{1,0}}^{\mathsf{cb}}\left(\chi_{1/2}^{\Vir}\chi_{\bm{1}}^{E_8} + \chi_{0}^{\Vir}\chi_{\bm{3875}}^{E_8} + \chi_{1/16}^{\Vir}\chi_{\bm{248}}^{E_8}\right) - \sZ_{\mathsf{1,1}}^{\mathsf{cb}}\left(-\chi_{1/2}^{\Vir}\chi_{\bm{1}}^{E_8} - \chi_{0}^{\Vir}\chi_{\bm{3875}}^{E_8} + \chi_{1/16}^{\Vir}\chi_{\bm{248}}^{E_8}\right) \right] ~, \label{eq:Z-F1-RNS}\\
\hspace{-0.15in}\sZ_{\sR,\sR}^{\sT_1} &= \frac{1}{2}\left[ \sZ_{\mathsf{1,0}}^{\mathsf{cb}}\left(\chi_{1/2}^{\Vir}\chi_{\bm{1}}^{E_8} + \chi_{0}^{\Vir}\chi_{\bm{3875}}^{E_8} + \chi_{1/16}^{\Vir}\chi_{\bm{248}}^{E_8}\right) + \sZ_{\mathsf{1,1}}^{\mathsf{cb}}\left(-\chi_{1/2}^{\Vir}\chi_{\bm{1}}^{E_8} - \chi_{0}^{\Vir}\chi_{\bm{3875}}^{E_8} + \chi_{1/16}^{\Vir}\chi_{\bm{248}}^{E_8}\right) \right. \nonumber\\
\hspace{-0.15in}&\qquad - \left. \sZ_{\mathsf{1,0}}^{\mathsf{cb}}\left(\chi_{1/2}^{\Vir}\chi_{\bm{1}}^{E_8} + \chi_{0}^{\Vir}\chi_{\bm{3875}}^{E_8} + \chi_{1/16}^{\Vir}\chi_{\bm{248}}^{E_8}\right) + \sZ_{\mathsf{1,1}}^{\mathsf{cb}}\left(-\chi_{1/2}^{\Vir}\chi_{\bm{1}}^{E_8} - \chi_{0}^{\Vir}\chi_{\bm{3875}}^{E_8} + \chi_{1/16}^{\Vir}\chi_{\bm{248}}^{E_8}\right) \right] ~. \label{eq:Z-F1-RR}
\end{align}
Another combination that we can consider is the tensor product,
\eqa{
  \cA_2 := \cbos_{2R'} \otimes (E_8)_2 \times \lambda ~. \label{eq:Theory-CA2}
}
This is a fermionic theory based on \eqref{eq:e8-level2-psi}. We consider the orbifold of this theory by the diagonal of the product of the half-period $\IZ_2^\hp$ acting on the compact boson and the fermion parity ($(-1)^{F}$) action of $(E_8)_2 \times \lambda$. Gauging $(-1)^{F}$ of a fermionic theory was introduced in Section \ref{sec:fermparity}. Following \eqref{eq:zpec}, we define the partition function of theory $\cA_2$ as
\eqa{
 &\bm{\sZ}^{\cA_2}_{p,\sfa;q,\sfb} &&:= \sZ_{\sfa,\sfb}(\cbos_{2R'})\times \bm{\sZ}_{p,\sfa;q,\sfb}^{(E_8)_2 \times \lambda} = \sZ_{\sfa,\sfb}(\cbos_{2R'})\times \sZ_{p+\sfa,q+\sfb}^{(E_8)_2 \times \lambda}  ~.
}
Then, the orbifold theory
\eqa{
  &\sT_2 &&:= \cA_{2}\orb\IZ_2 = \big(\cbos_{2R'} \otimes (E_8)_2 \times \lambda\big)\orb\mathsf{diag}\big(\IZ_2^\hp \times (-1)^{F}\big) ~, \label{eq:Theory-CF2}
}
has partition function,
\eqa{
  &\sZ^{\sT_2}_{p,q} &&= \frac{1}{2}\sum_{\sfa,\sfb\in\{0,1\}}\bm{\sZ}^{\cA_2}_{p,\sfa;q,\sfb} = \frac{1}{2}\sum_{\sfa,\sfb\in\{0,1\}}\left( \sZ_{\sfa,\sfb}(\cbos_{2R'})\times \sZ_{p+\sfa,q+\sfb}^{(E_8)_2 \times \lambda} \right) ~. \label{eq:Z-CF2-pq}
}
As before, we define
\eqa{
  \sZ^{\mathsf{cb}'}_{\mathsf{a,b}} := \sZ_{\mathsf{a,b}}(\cbos_{2R'}; \IZ_{2}^{\hp}) ~.
}
Evaluating \eqref{eq:Z-CF2-pq} explicitly, we find:
\eqa{
& \sZ_{\sN\sS,\sN\sS}^{\sT_2} 
%
 &&= \frac{1}{2}\sZ^{\mathsf{cb}'}_{\mathsf{0},\mathsf{0}}\left(\chi_{0}^{\Vir}\chi_{\bm{1}}^{E_8} + \chi_{1/2}^{\Vir}\chi_{\bm{1}}^{E_8} + \chi_{0}^{\Vir}\chi_{\bm{3875}}^{E_8} + \chi_{1/2}^{\Vir}\chi_{\bm{3875}}^{E_8} \right) \nonumber\\
& &&\quad + \frac{1}{2}\sZ^{\mathsf{cb}'}_{\mathsf{0},\mathsf{1}}\left(\chi_{0}^{\Vir}\chi_{\bm{1}}^{E_8} - \chi_{1/2}^{\Vir}\chi_{\bm{1}}^{E_8} - \chi_{0}^{\Vir}\chi_{\bm{3875}}^{E_8} + \chi_{1/2}^{\Vir}\chi_{\bm{3875}}^{E_8} \right) + \sZ_{\mathsf{1},\mathsf{0}}^{\mathsf{cb}'}\chi_{1/16}^{\Vir}\chi_{\bm{248}}^{E_8} ~, \label{eq:Z-F2-NSR}\\
& \sZ_{\sN\sS,\sR}^{\sT_2} 
%
 &&= \frac{1}{2}\sZ^{\mathsf{cb}'}_{\mathsf{0},\mathsf{1}}\left( \chi_{0}^{\Vir}\chi_{\bm{1}}^{E_8} + \chi_{1/2}^{\Vir}\chi_{\bm{1}}^{E_8} + \chi_{0}^{\Vir}\chi_{\bm{3875}}^{E_8} + \chi_{1/2}^{\Vir}\chi_{\bm{3875}}^{E_8} \right) \nonumber\\
& &&\quad + \frac{1}{2}\sZ^{\mathsf{cb}'}_{\mathsf{0},\mathsf{0}}\left( \chi_{0}^{\Vir}\chi_{\bm{1}}^{E_8} - \chi_{1/2}^{\Vir}\chi_{\bm{1}}^{E_8} - \chi_{0}^{\Vir}\chi_{\bm{3875}}^{E_8} + \chi_{1/2}^{\Vir}\chi_{\bm{3875}}^{E_8} \right) + \sZ_{\mathsf{1},\mathsf{1}}^{\mathsf{cb}'}\chi_{1/16}^{\Vir}\chi_{\bm{248}}^{E_8} ~, \label{eq:Z-F2-NSNS}\\
& \sZ_{\sR,\sN\sS}^{\sT_2} 
%
  &&= \frac{1}{2}\sZ^{\mathsf{cb}'}_{\mathsf{1},\mathsf{0}}\left(\chi_{0}^{\Vir}\chi_{\bm{1}}^{E_8} + \chi_{1/2}^{\Vir}\chi_{\bm{1}}^{E_8} + \chi_{0}^{\Vir}\chi_{\bm{3875}}^{E_8} + \chi_{1/2}^{\Vir}\chi_{\bm{3875}}^{E_8} \right) \nonumber\\
 & &&\quad + \frac{1}{2}\sZ^{\mathsf{cb}'}_{\mathsf{1},\mathsf{1}}\left(\chi_{0}^{\Vir}\chi_{\bm{1}}^{E_8} - \chi_{1/2}^{\Vir}\chi_{\bm{1}}^{E_8} - \chi_{0}^{\Vir}\chi_{\bm{3875}}^{E_8} + \chi_{1/2}^{\Vir}\chi_{\bm{3875}}^{E_8}\right) + \sZ_{\mathsf{0},\mathsf{0}}^{\mathsf{cb}'}\chi_{1/16}^{\Vir} \chi_{\bm{248}}^{E_8} ~ \label{eq:Z-F2-RNS}\\
& \sZ_{\sR,\sR}^{\sT_2} 
%
&&= \frac{1}{2}\sZ^{\mathsf{cb}'}_{\mathsf{1},\mathsf{1}}\left( \chi_{0}^{\Vir}\chi_{\bm{1}}^{E_8} + \chi_{1/2}^{\Vir}\chi_{\bm{1}}^{E_8} + \chi_{0}^{\Vir}\chi_{\bm{3875}}^{E_8} + \chi_{1/2}^{\Vir}\chi_{\bm{3875}}^{E_8}\right) \nonumber\\
& &&\quad + \frac{1}{2}\sZ^{\mathsf{cb}'}_{\mathsf{1},\mathsf{0}}\left( \chi_{0}^{\Vir}\chi_{\bm{1}}^{E_8} - \chi_{1/2}^{\Vir}\chi_{\bm{1}}^{E_8} - \chi_{0}^{\Vir}\chi_{\bm{3875}}^{E_8} + \chi_{1/2}^{\Vir}\chi_{\bm{3875}}^{E_8}\right) + \sZ_{\mathsf{0},\mathsf{1}}^{\mathsf{cb}'} \chi_{1/16}^{\Vir}\chi_{\bm{248}}^{E_8} \label{eq:Z-F2-RR} ~.
}
In the next section, after a slight but essential digression, we will show that $\sT_1$ and $\sT_2$ are T-dual to each other for $R' = \frac{1}{2R}$. While one can prove this by comparing \eqref{eq:Z-F1-NSNS}--\eqref{eq:Z-F1-RR} with \eqref{eq:Z-F2-NSNS}--\eqref{eq:Z-F2-RR}, doing so involves a tedious change of variables. We opt for a more conceptual explanation based on the methods of Section \ref{sec:modernz2orbifoldintro}.

\section{Fermionizations and T-Dualities}\label{sec:generaltduality}

\subsection{T-Duality of $\mathsf{T}_1$ and $\mathsf{T}_2$}
We first demonstrate the T-duality of a pair of 2d spin CFTs using the fermionization of Section \ref{sec:modernz2orbifoldintro}. Let $\cT_{\sA}$ be a left-moving bosonic CFT that admits a non-anomalous $\bZ_2^{(\sA)}$ symmetry, with $\sZ_{2}^{(\sA)}$-twisted partition functions denoted by $\sZ^{\cT_\sA}_{\mathsf{i},\mathsf{j}}$ where $\mathsf{i}, \mathsf{j} \in \{ \mathsf{0}, \mathsf{1}\}$. 
We can fermionize $\cT_\sA$ to obtain a theory $\cT_\sF := \big(\cT_{\sA}\times\Arf\big)/\mathsf{diag}\big(\IZ_2^{\sA}\times \IZ_2^\Arf\big)$ with partition functions \eqref{eq:fermNS-NS}--\eqref{eq:fermR-R} which read\footnote{We denote the fermionizations of $\cT_\sA$ by $\cT_\sF := \sF(\cT_\sA)$, and $\cT_{\sF'} := \sF'(\cT_\sA) = \sF(\cT_\sA)\times\Arf$.}
\eqa{
& \sZ^{\cT_\sF}_{\sN\sS,\sN\sS} &&= \sZ_\sS^{\cT_\sA} + \sZ_\sV^{\cT_\sA} = \frac{1}{2}\left( \sZ_{\mathsf{0},\mathsf{0}}^{\cT_\sA} + \sZ_{\mathsf{0},\mathsf{1}}^{\cT_\sA} +  \sZ_{\mathsf{1},\mathsf{0}}^{\cT_\sA} - \sZ_{\mathsf{1},\mathsf{1}}^{\cT_\sA}\right) ~, \label{eq:Z-TF-NSNS}\\
& \sZ^{\cT_\sF}_{\sN\sS,\sR} &&= \sZ_\sS^{\cT_\sA} - \sZ_\sV^{\cT_\sA} = \frac{1}{2}\left( \sZ_{\mathsf{0},\mathsf{0}}^{\cT_\sA} + \sZ_{\mathsf{0},\mathsf{1}}^{\cT_\sA} -  \sZ_{\mathsf{1},\mathsf{0}}^{\cT_\sA} + \sZ_{\mathsf{1},\mathsf{1}}^{\cT_\sA}\right) ~, \label{eq:Z-TF-NSR}\\
& \sZ^{\cT_\sF}_{\sR,\sN\sS} &&= \sZ_\sU^{\cT_\sA} + \sZ_\sT^{\cT_\sA} = \frac{1}{2}\left( \sZ_{\mathsf{1},\mathsf{0}}^{\cT_\sA} + \sZ_{\mathsf{1},\mathsf{1}}^{\cT_\sA} + \sZ_{\mathsf{0},\mathsf{0}}^{\cT_\sA} - \sZ_{\mathsf{0},\mathsf{1}}^{\cT_\sA}\right) ~, \label{eq:Z-TF-RNS}\\
& \sZ^{\cT_\sF}_{\sR,\sR} &&= \sZ_\sU^{\cT_\sA} - \sZ_\sT^{\cT_\sA} = \frac{1}{2}\left( \sZ_{\mathsf{1},\mathsf{0}}^{\cT_\sA} + \sZ_{\mathsf{1},\mathsf{1}}^{\cT_\sA} - \sZ_{\mathsf{0},\mathsf{0}}^{\cT_\sA} + \sZ_{\mathsf{0},\mathsf{1}}^{\cT_\sA}\right) ~. \label{eq:Z-TF-RR}
}

\subsection*{Construction of theory $\cX_1$}
Consider the theory,
\eqa{
& (\text{compact boson at radius $2R$} ) \otimes \cT_\sA ~,
}
which we will abbreviate by `$\cbos \otimes \cT_\sA$'. This theory has a $\bZ_2^{\hp} \times \bZ_2^{(\sA)}$ symmetry, with the first factor corresponding to the half-period shift of the compact boson and the second factor the $\bZ_2$ symmetry of $\cT_\sA$. We are interested in the diagonal $\bZ_2$ subgroup, with respect to which we can construct the twisted and untwisted sectors. The $\sS$, $\sT$, $\sU$ and $\sV$ sectors of $\cbos \otimes \cT_\sA$ have the following partition functions:
\eqa{
& \sZ^{\cbos\otimes\cT_\sA}_{\sS} &&= \frac{1}{2}(\sZ^{\cbos}_{\mathsf{0,0}}\sZ^{\cT_\sA}_{\mathsf{0,0}} + \sZ^{\cbos}_{\mathsf{0,1}}\sZ^{\cT_\sA}_{\mathsf{0,1}}) ~,\\
& \sZ^{\cbos\otimes\cT_\sA}_{\sT} &&= \frac{1}{2}(\sZ^{\cbos}_{\mathsf{0,0}}\sZ^{\cT_\sA}_{\mathsf{0,0}} - \sZ^{\cbos}_{\mathsf{0,1}}\sZ^{\cT_\sA}_{\mathsf{0,1}}) ~,\\
& \sZ^{\cbos\otimes\cT_\sA}_{\sU} &&= \frac{1}{2}(\sZ^{\cbos}_{\mathsf{1,0}}\sZ^{\cT_\sA}_{\mathsf{1,0}} + \sZ^{\cbos}_{\mathsf{1,1}}\sZ^{\cT_\sA}_{\mathsf{1,1}}) ~,\\
& \sZ^{\cbos\otimes\cT_\sA}_{\sV} &&= \frac{1}{2}(\sZ^{\cbos}_{\mathsf{1,0}}\sZ^{\cT_\sA}_{\mathsf{1,0}} - \sZ^{\cbos}_{\mathsf{1,1}}\sZ^{\cT_\sA}_{\mathsf{1,1}}) ~.
}
The corresponding fermionized theory,
\eqa{
  \cX_1 &:= \sF(\cbos_{2R}\otimes\cT_\sA) = \big(\cbos_{2R} \otimes \cT_{\sA} \otimes \Arf)\orb\mathsf{diag}\big( \IZ_2^\hp \times \IZ_2^{(\sA)} \times \IZ_2^\Arf\big) ~, \label{eq:defn-X1}
}
has the following partition functions:
\eqa{
& \sZ^{\cX_1}_{\sN\sS,\sN\sS} &&= \sZ^{\cbos\otimes\cT_\sA}_{\sS} + \sZ^{\cbos\otimes\cT_\sA}_{\sV} = \frac{1}{2}\left(\sZ^{\cbos}_{\mathsf{0,0}}\sZ^{\cT_\sA}_{\mathsf{0,0}} + \sZ^{\cbos}_{\mathsf{0,1}}\sZ^{\cT_\sA}_{\mathsf{0,1}} + \sZ^{\cbos}_{\mathsf{1,0}}\sZ^{\cT_\sA}_{\mathsf{1,0}} - \sZ^{\cbos}_{\mathsf{1,1}}\sZ^{\cT_\sA}_{\mathsf{1,1}}\right) ~, \label{eq:X1-STUV-NSNS}\\
& \sZ^{\cX_1}_{\sN\sS,\sR} &&= \sZ^{\cbos\otimes\cT_\sA}_{\sS} - \sZ^{\cbos\otimes\cT_\sA}_{\sV} =  \frac{1}{2}\left(\sZ^{\cbos}_{\mathsf{0,0}}\sZ^{\cT_\sA}_{\mathsf{0,0}} + \sZ^{\cbos}_{\mathsf{0,1}}\sZ^{\cT_\sA}_{\mathsf{0,1}} - \sZ^{\cbos}_{\mathsf{1,0}}\sZ^{\cT_\sA}_{\mathsf{1,0}} + \sZ^{\cbos}_{\mathsf{1,1}}\sZ^{\cT_\sA}_{\mathsf{1,1}}\right) ~, \label{eq:X1-STUV-NSR}\\
& \sZ^{\cX_1}_{\sR,\sN\sS} &&= \sZ^{\cbos\otimes\cT_\sA}_{\sU} + \sZ^{\cbos\otimes\cT_\sA}_{\sT} = \frac{1}{2}\left(\sZ^{\cbos}_{\mathsf{1,0}}\sZ^{\cT_\sA}_{\mathsf{1,0}} + \sZ^{\cbos}_{\mathsf{1,1}}\sZ^{\cT_\sA}_{\mathsf{1,1}} + \sZ^{\cbos}_{\mathsf{0,0}}\sZ^{\cT_\sA}_{\mathsf{0,0}} - \sZ^{\cbos}_{\mathsf{0,1}}\sZ^{\cT_\sA}_{\mathsf{0,1}} \right)~, \label{eq:X1-STUV-RNS}\\
& \sZ^{\cX_1}_{\sR,\sR} &&= \sZ^{\cbos\otimes\cT_\sA}_{\sU} - \sZ^{\cbos\otimes\cT_\sA}_{\sT} = \frac{1}{2}\left(\sZ^{\cbos}_{\mathsf{1,0}}\sZ^{\cT_\sA}_{\mathsf{1,0}} + \sZ^{\cbos}_{\mathsf{1,1}}\sZ^{\cT_\sA}_{\mathsf{1,1}} - \sZ^{\cbos}_{\mathsf{0,0}}\sZ^{\cT_\sA}_{\mathsf{0,0}} + \sZ^{\cbos}_{\mathsf{0,1}}\sZ^{\cT_\sA}_{\mathsf{0,1}} \right) ~.\label{eq:X1-STUV-RR}
}
In terms of the $\sS$, $\sT$, $\sU$, $\sV$ sectors of $\cbos$ and $\cT_{\sA}$, they reduce to:
\eqa{
& \sZ^{\cX_1}_{\sN\sS,\sN\sS} &&= \sZ^{\cbos}_{\sS} \sZ^{\cT_\sA}_{\sS} + \sZ^{\cbos}_{\sT} \sZ^{\cT_\sA}_{\sT} + \sZ^{\cbos}_{\sV} \sZ^{\cT_\sA}_{\sU} + \sZ^{\cbos}_{\sU} \sZ^{\cT_\sA}_{\sV} ~, \label{eq:X1-STUV-NSNS-final}\\
& \sZ^{\cX_1}_{\sN\sS,\sR} &&= \sZ^{\cbos}_{\sS} \sZ^{\cT_\sA}_{\sS} + \sZ^{\cbos}_{\sT} \sZ^{\cT_\sA}_{\sT} - \sZ^{\cbos}_{\sV} \sZ^{\cT_\sA}_{\sU} - \sZ^{\cbos}_{\sU} \sZ^{\cT_\sA}_{\sV} ~, \label{eq:X1-STUV-NSR-final}\\
& \sZ^{\cX_1}_{\sR,\sN\sS} &&= \sZ^{\cbos}_{\sT} \sZ^{\cT_\sA}_{\sS} + \sZ^{\cbos}_{\sS} \sZ^{\cT_\sA}_{\sT} + \sZ^{\cbos}_{\sU} \sZ^{\cT_\sA}_{\sU} + \sZ^{\cbos}_{\sV} \sZ^{\cT_\sA}_{\sV} ~, \label{eq:X1-STUV-RNS-final}\\
& \sZ^{\cX_1}_{\sR,\sR} &&= -\sZ^{\cbos}_{\sT} \sZ^{\cT_\sA}_{\sS} - \sZ^{\cbos}_{\sS} \sZ^{\cT_\sA}_{\sT} + \sZ^{\cbos}_{\sU} \sZ^{\cT_\sA}_{\sU} + \sZ^{\cbos}_{\sV} \sZ^{\cT_\sA}_{\sV} ~. \label{eq:X1-STUV-RR-final}
}

\subsection*{Construction of theory $\cX_2$}
The theory $\cX_2$ is defined as,
\eqa{
  \cX_2 := \big(\cbos_{2R'} \otimes \cT_{\sF'} \big)\orb \mathsf{diag}\big( \IZ_2^\hp \times (-1)^{F}\big) ~,
}
where $\cT_{\sF'} = \sF(\cT_{\sA}) \otimes \Arf = (\cT_{\sA} \otimes \Arf)\orb \mathsf{diag}(\IZ_2^{(\sA)}\times \IZ_2^\Arf) \otimes \Arf$. Following Section \ref{sec:fermparity}, the partition function of this orbifold theory is
\eqa{
& \sZ_{p,q}^{\cX_2} &&:= \frac{1}{2}\sum_{a,b \in \{0, 1\}} \left( \sZ^{\cbos'}_{\sfa,\sfb} \times \bm{\sZ}^{\cT_{\sF'}}_{p,\sfa; q,\sfb}\right) ~. \label{eq:Z-cX2-pq}
}
Evaluating \eqref{eq:Z-cX2-pq} explicitly, we find:
\eqa{
& \sZ^{\cX_2}_{\sN\sS,\sN\sS} &&= \frac{1}{2}\left( \sZ_{\mathsf{0,0}}^{\cbos'} \sZ_{\sN\sS,\sN\sS}^{\cT_\sF} + \sZ_{\mathsf{0,1}}^{\cbos'} \sZ_{\sN\sS,\sR}^{\cT_\sF} + \sZ_{\mathsf{1,0}}^{\cbos'} \sZ_{\sR,\sN\sS}^{\cT_\sF} - \sZ_{\mathsf{1,1}}^{\cbos'} \sZ_{\sR,\sR}^{\cT_\sF}\right) ~, \label{eq:Z-cX2'-NSNS}\\
& \sZ^{\cX_2}_{\sN\sS,\sR} &&= \frac{1}{2}\left( \sZ_{\mathsf{0,0}}^{\cbos'} \sZ_{\sN\sS,\sR}^{\cT_\sF} + \sZ_{\mathsf{0,1}}^{\cbos'} \sZ_{\sN\sS,\sN\sS}^{\cT_\sF} - \sZ_{\mathsf{1,0}}^{\cbos'} \sZ_{\sR,\sR}^{\cT_\sF} + \sZ_{\mathsf{1,1}}^{\cbos'} \sZ_{\sR,\sN\sS}^{\cT_\sF}\right) ~, \label{eq:Z-cX2'-NSR}\\
& \sZ^{\cX_2}_{\sR,\sN\sS} &&= \frac{1}{2}\left( \sZ_{\mathsf{0,0}}^{\cbos'} \sZ_{\sR,\sN\sS}^{\cT_\sF} - \sZ_{\mathsf{0,1}}^{\cbos'} \sZ_{\sR,\sR}^{\cT_\sF} + \sZ_{\mathsf{1,0}}^{\cbos'} \sZ_{\sN\sS,\sN\sS}^{\cT_\sF} + \sZ_{\mathsf{1,1}}^{\cbos'} \sZ_{\sN\sS,\sR}^{\cT_\sF}\right) ~, \label{eq:Z-cX2'-RNS}\\
& \sZ^{\cX_2}_{\sR,\sR} &&= \frac{1}{2}\left(-\sZ_{\mathsf{0,0}}^{\cbos'} \sZ_{\sR,\sR}^{\cT_\sF} + \sZ_{\mathsf{0,1}}^{\cbos'} \sZ_{\sR,\sN\sS}^{\cT_\sF} + \sZ_{\mathsf{1,0}}^{\cbos'} \sZ_{\sN\sS,\sR}^{\cT_\sF} + \sZ_{\mathsf{1,1}}^{\cbos'} \sZ_{\sN\sS,\sN\sS}^{\cT_\sF}\right) ~. \label{eq:Z-cX2'-RR}
}
Using \eqref{eq:Z-TF-NSNS}--\eqref{eq:Z-TF-RR}, these can be written as
\eqa{
& \sZ^{\cX_2}_{\sN\sS,\sN\sS} &&= \tfrac{1}{4}(\sZ^{\cbos'}_{\mathsf{0,0}} + \sZ^{\cbos'}_{\mathsf{0,1}} + \sZ^{\cbos'}_{\mathsf{1,0}} 
+ \sZ^{\cbos'}_{\mathsf{1,1}})\sZ^{\cT_\sA}_{\mathsf{0,0}}  +  \tfrac{1}{4}(\sZ^{\cbos'}_{\mathsf{0,0}} + \sZ^{\cbos'}_{\mathsf{0,1}} - \sZ^{\cbos'}_{\mathsf{1,0}} - \sZ^{\cbos'}_{\mathsf{1,1}})\sZ^{\cT_\sA}_{\mathsf{0,1}} \nonumber\\
& &&\quad +  \tfrac{1}{4}(\sZ^{\cbos'}_{\mathsf{0,0}} - \sZ^{\cbos'}_{\mathsf{0,1}} + \sZ^{\cbos'}_{\mathsf{1,0}} - \sZ^{\cbos'}_{\mathsf{1,1}})\sZ^{\cT_\sA}_{\mathsf{1,0}} 
+  \tfrac{1}{4}(-\sZ^{\cbos'}_{\mathsf{0,0}} + \sZ^{\cbos'}_{\mathsf{0,1}} + \sZ^{\cbos'}_{\mathsf{1,0}} - \sZ^{\cbos'}_{\mathsf{1,1}})\sZ^{\cT_\sA}_{\mathsf{1,1}} ~, \label{eq:X2-STUV-NSNS}\\
& \sZ^{\cX_2}_{\sN\sS,\sR} &&= \tfrac{1}{4}(\sZ^{\cbos'}_{\mathsf{0,0}} + \sZ^{\cbos'}_{\mathsf{0,1}} + \sZ^{\cbos'}_{\mathsf{1,0}} + \sZ^{\cbos'}_{\mathsf{1,1}})\sZ^{\cT_\sA}_{\mathsf{0,0}}
 +  \tfrac{1}{4}(\sZ^{\cbos'}_{\mathsf{0,0}} + \sZ^{\cbos'}_{\mathsf{0,1}} - \sZ^{\cbos'}_{\mathsf{1,0}} - \sZ^{\cbos'}_{\mathsf{1,1}})\sZ^{\cT_\sA}_{\mathsf{0,1}} \nonumber\\
& &&\quad +  \tfrac{1}{4}(-\sZ^{\cbos'}_{\mathsf{0,0}} + \sZ^{\cbos'}_{\mathsf{0,1}} - \sZ^{\cbos'}_{\mathsf{1,0}} + \sZ^{\cbos'}_{\mathsf{1,1}})\sZ^{\cT_\sA}_{\mathsf{1,0}} 
+  \tfrac{1}{4}(\sZ^{\cbos'}_{\mathsf{0,0}} - \sZ^{\cbos'}_{\mathsf{0,1}} - \sZ^{\cbos'}_{\mathsf{1,0}} + \sZ^{\cbos'}_{\mathsf{1,1}})\sZ^{\cT_\sA}_{\mathsf{1,1}} ~, \label{eq:X2-STUV-NSR}\\
& \sZ^{\cX_2}_{\sR,\sN\sS} &&= \tfrac{1}{4}(\sZ^{\cbos'}_{\mathsf{0,0}} + \sZ^{\cbos'}_{\mathsf{0,1}} + \sZ^{\cbos'}_{\mathsf{1,0}} + \sZ^{\cbos'}_{\mathsf{1,1}})\sZ^{\cT_\sA}_{\mathsf{0,0}}
 +  \tfrac{1}{4}(-\sZ^{\cbos'}_{\mathsf{0,0}} - \sZ^{\cbos'}_{\mathsf{0,1}} + \sZ^{\cbos'}_{\mathsf{1,0}} + \sZ^{\cbos'}_{\mathsf{1,1}})\sZ^{\cT_\sA}_{\mathsf{0,1}} \nonumber\\
& &&\quad +  \tfrac{1}{4}(\sZ^{\cbos'}_{\mathsf{0,0}} - \sZ^{\cbos'}_{\mathsf{0,1}} + \sZ^{\cbos'}_{\mathsf{1,0}} - \sZ^{\cbos'}_{\mathsf{1,1}})\sZ^{\cT_\sA}_{\mathsf{1,0}} 
+  \tfrac{1}{4}(\sZ^{\cbos'}_{\mathsf{0,0}} - \sZ^{\cbos'}_{\mathsf{0,1}} - \sZ^{\cbos'}_{\mathsf{1,0}} + \sZ^{\cbos'}_{\mathsf{1,1}})\sZ^{\cT_\sA}_{\mathsf{1,1}} ~, \label{eq:X2-STUV-RNS}\\
& \sZ^{\cX_2}_{\sR,\sR} &&= \tfrac{1}{4}(\sZ^{\cbos'}_{\mathsf{0,0}} + \sZ^{\cbos'}_{\mathsf{0,1}} + \sZ^{\cbos'}_{\mathsf{1,0}} + \sZ^{\cbos'}_{\mathsf{1,1}})\sZ^{\cT_\sA}_{\mathsf{0,0}}
 +  \tfrac{1}{4}(-\sZ^{\cbos'}_{\mathsf{0,0}} - \sZ^{\cbos'}_{\mathsf{0,1}} + \sZ^{\cbos'}_{\mathsf{1,0}} + \sZ^{\cbos'}_{\mathsf{1,1}})\sZ^{\cT_\sA}_{\mathsf{0,1}} \nonumber\\
& &&\quad +  \tfrac{1}{4}(-\sZ^{\cbos'}_{\mathsf{0,0}} + \sZ^{\cbos'}_{\mathsf{0,1}} - \sZ^{\cbos'}_{\mathsf{1,0}} + \sZ^{\cbos'}_{\mathsf{1,1}})\sZ^{\cT_\sA}_{\mathsf{1,0}} 
+  \tfrac{1}{4}(-\sZ^{\cbos'}_{\mathsf{0,0}} + \sZ^{\cbos'}_{\mathsf{0,1}} + \sZ^{\cbos'}_{\mathsf{1,0}} - \sZ^{\cbos'}_{\mathsf{1,1}})\sZ^{\cT_\sA}_{\mathsf{1,1}} ~. \label{eq:X2-STUV-RR}
}
In terms of $\sS, \sT, \sU, \sV$ sectors of $\cbos'$ and $\cT_\sA$, they reduce to:
\eqa{
& \sZ^{\cX_2}_{\sN\sS,\sN\sS} &&= \sZ^{\cbos'}_{\sS}\sZ^{\cT_\sA}_{\sS} +  \sZ^{\cbos'}_{\sU}\sZ^{\cT_\sA}_{\sT} + \sZ^{\cbos'}_{\sV}\sZ^{\cT_\sA}_{\sU} + \sZ^{\cbos'}_{\sT}\sZ^{\cT_\sA}_{\sV}  ~, \label{eq:X2-STUV-NSNS-final}\\
& \sZ^{\cX_2}_{\sN\sS,\sR} &&= \sZ^{\cbos'}_{\sS}\sZ^{\cT_\sA}_{\sS} +  \sZ^{\cbos'}_{\sU}\sZ^{\cT_\sA}_{\sT} - \sZ^{\cbos'}_{\sV}\sZ^{\cT_\sA}_{\sU} - \sZ^{\cbos'}_{\sT}\sZ^{\cT_\sA}_{\sV}  ~, \label{eq:X2-STUV-NSR-final}\\
& \sZ^{\cX_2}_{\sR,\sN\sS} &&= \sZ^{\cbos'}_{\sU}\sZ^{\cT_\sA}_{\sS} +  \sZ^{\cbos'}_{\sS}\sZ^{\cT_\sA}_{\sT} + \sZ^{\cbos'}_{\sT}\sZ^{\cT_\sA}_{\sU} + \sZ^{\cbos'}_{\sV}\sZ^{\cT_\sA}_{\sV}  ~, \label{eq:X2-STUV-RNS-final}\\
& \sZ^{\cX_2}_{\sR,\sR} &&= \sZ^{\cbos'}_{\sU}\sZ^{\cT_\sA}_{\sS} +  \sZ^{\cbos'}_{\sS}\sZ^{\cT_\sA}_{\sT} - \sZ^{\cbos'}_{\sT}\sZ^{\cT_\sA}_{\sU} - \sZ^{\cbos'}_{\sV}\sZ^{\cT_\sA}_{\sV}  ~. \label{eq:X2-STUV-RR-final}
}
\subsection*{T-Duality of $\cX_1$ and $\cX_2$}
Note that the theory $\cT_\sA$ does not participate in the T-duality and serves merely as a spectator. For $R' = \frac{1}{2R}$, 
\eqa{
\sZ^{\cbos}_{\sS} &= \sZ^{\cbos'}_{\sS} ~, \quad
\sZ^{\cbos}_{\sU} &= \sZ^{\cbos'}_{\sT} ~, \quad 
\sZ^{\cbos}_{\sT} &= \sZ^{\cbos'}_{\sU} ~, \quad
\sZ^{\cbos}_{\sV} &&= \sZ^{\cbos'}_{\sV} ~. \label{eq:cb-cb'-tduality}
}
So the $\sS$ and $\sV$ sectors of the compact boson are mapped into themselves, whereas the $\sT$ and $\sU$ sectors are exchanged. Using \eqref{eq:cb-cb'-tduality} (with $R' = \frac{1}{2R}$) in \eqref{eq:X1-STUV-NSNS-final}--\eqref{eq:X1-STUV-RR-final} and \eqref{eq:X2-STUV-NSNS-final}--\eqref{eq:X2-STUV-RR-final}, it is easy to verify that
\eqa{
& \sZ^{\cX_1}_{\sN\sS,\sN\sS} &=  \sZ^{\cX_2}_{\sN\sS,\sN\sS} ~, \quad
\sZ^{\cX_1}_{\sN\sS,\sR} &=  \sZ^{\cX_2}_{\sN\sS,\sR} ~, \quad
\sZ^{\cX_1}_{\sR,\sN\sS} &=  \sZ^{\cX_2}_{\sR,\sN\sS} ~, \quad 
\sZ^{\cX_1}_{\sR,\sR} &= - \sZ^{\cX_2}_{\sR,\sR} ~.
}
Therefore, the theories
\eqa{
& \cX_1 := \big[\cbos_{2R} \otimes \cT_\sA \otimes \Arf\big]\orb\mathsf{diag}\big(\IZ_2^\hp \times \IZ_2^{(\sA)} \times \IZ_2^\Arf\big) ~,
}
and
\eqa{
& \cX_2 := \big[\cbos_{2R'} \otimes \cT_{\sF'}\big]\orb\mathsf{diag}\big(\IZ_2^\hp \times (-1)^{F}\big) ~,
}
are T-dual to each other for $R' = \frac{1}{2R}$. We emphasize the crucial role of the additional $\Arf$ theory in the definition of the theory $\cX_2$ (we used $\cT_{\sF'}$, and not $\cT_{\sF}$ corresponding to a given $\cT_{\sA}$). Whenever $\sZ^{\cT_\sA}_{\sT}$ and  $\sZ^{\cT_\sA}_{\sU}$ are equal, the fermionizations $\cT_{\sF}$ and $\cT_{\sF'}$ coincide and this additional stacking with the $\Arf$ theory is unnecessary. 

\subsection*{T-Duality of $\sT_{1}$ and $\sT_{2}$}
For $\cT_{\sA} = (E_8)_1 \times (E_8)_1$, the fermionizations $\cT_{\sF'}$ and $\cT_{\sF}$ coincide because the $\sT$ and $\sU$ sectors are identical, see \eqref{eq:E8byE8-T}, \eqref{eq:E8byE8-U}. Applying the general result above, we directly conclude that the theories,
\eqa{
  &\sT_1 &&:= \left[\cbos_{2R} \otimes (E_8)_1 \times (E_8)_1 \otimes \Arf\right]\orb\mathsf{diag}\big( \IZ_2^\hp \times \IZ_2^\sigma \times \IZ_2^\Arf \big) ~,
}
and 
\eqa{
  &\sT_2 &&:= \left[\cbos_{2R'} \otimes (E_8)_2 \times \lambda\right]\orb\mathsf{diag}\big( \IZ_2^\hp \times (-1)^{F} \big) ~,
}
are $T$-dual to each other for $R' = \frac{1}{2R}$. Both $\sT_1$ and $\sT_2$ have central charge $(c_L, c_R)=(17,1)$.

\subsection{Fermionization and Spacetime Spin Structure}\label{subsec:spacetimespinstructure}

An important feature of fermionizing the compact boson by the $\Arf$ theory is that it endows the spacetime fermions with an antiperiodic spin structure around $S^{1}_{R}$. To see this, it is instructive to consider the following 2d $\cN=(0,1)$ SCFT with central charge $(17,\frac{3}{2})$ as modeling the internal degrees of freedom of the $9d$ non-supersymmetric heterotic string,
\eqas{
  \mathsf{T}_{\mathsf{int}}^{(1)} := (E_8)_1 \times (E_8)_1 \otimes \big[ \mathsf{cb}_{2R} \times \mathsf{Arf}\big]\orb\IZ_2 \otimes \wt{\psi} ~,
}
and determine the spacetime fermionic spectrum in lightcone gauge. Including the spacetime degrees of freedom, the GSO-projected partition function in the $\sR$ sector is
\eqa{
\hspace{-0.1in}\left.\sZ_{\sR}^{(1)}\right|_{\mathsf{GSO}} &\sim \frac{(\mathsf{Im\,}\tau)^{-7/2}}{|\eta(\tau)|^{14}}\overline{\bigg(\frac{\vartheta_2}{\eta}\bigg)^4}\big(\chi_{0}^{\Vir}\chi_{\bm{1}}^{E_8} + \chi_{1/2}^{\Vir}\chi_{\bm{3875}}^{E_8} + \chi_{1/16}^{\Vir}\chi_{\bm{248}}^{E_8}\big)\big(\sZ_{\sU}(\cbos_{2R}; \IZ_{2}^{\hp}) + \sZ_{\sT}(\cbos_{2R}; \IZ_{2}^{\hp})\big) ~,
}
up to a nonzero numerical coefficient that we omit for brevity. Recall that the Kaluza-Klein (KK) mass squared contribution of a scalar field compactified on a circle of radius $R$ is of the form $M^2 \sim m^2/R^2$ if the scalar obeys periodic boundary conditions and is $M^2 \sim \left(m + \frac{1}{2}\right)^2/R^2$ if it obeys antiperiodic boundary conditions. From \eqref{eq:cbos-2R-T} and \eqref{eq:cbos-2R-U}, the total scaling dimension contributions from the $\sT$ and $\sU$ sectors are
\eqas{
  \left.\begin{split}
    \sU &\quad:\quad \frac{m^2}{2R^2} + 2 n^2 R^2 ~,\\
    \sT &\quad:\quad \frac{\left(m+\frac{1}{2}\right)^2}{2R^2} + 2\left(n+\tfrac{1}{2}\right)^2 R^2 ~,
  \end{split}\right\} \quad  m, n \in \bZ ~.
}
From the nine-dimensional viewpoint, the Kaluza-Klein modes of the 10d gravitino have zero winding numbers. Between $\sU$ and $\sT$, only sector $\sT$ contributes such states for which the mass squared contribution reflects the spacetime antiperiodic boundary condition around $S^{1}_{R}$.

If we repeat the preceding analysis for the following internal theory with central charge $(17,\frac{3}{2})$:
\eqas{
  \mathsf{T}_{\mathsf{int}}^{(2)} := (E_8)_1 \times (E_8)_1 \otimes \mathsf{cb}_{R} \otimes \wt{\psi} ~,
}
we find (again up to an overall nonzero numerical coefficient),
\eqa{
  &\left.\sZ_{\sR}^{(2)}\right|_{\mathsf{GSO}} &&\sim \frac{(\mathsf{Im\,}\tau)^{-7/2}}{|\eta(\tau)|^{14}}\overline{\bigg(\frac{\vartheta_2}{\eta}\bigg)^4}\big(\chi_{0}^{\Vir}\chi_{\bm{1}}^{E_8} + \chi_{1/2}^{\Vir}\chi_{\bm{3875}}^{E_8} + \chi_{1/16}^{\Vir}\chi_{\bm{248}}^{E_8}\big)\sZ(\cbos_{R}) ~. 
}
From \eqref{eq:pf-cboson}, the total scaling dimension of a typical state of $\cbos_R$ with zero winding number is $\frac{m^2}{2R^2}$ for $m \in \IZ$, reflecting the spacetime periodic boundary condition around $S_{R}^1$.

To summarize, the spacetime spin structure -- which manifests in this case (in the $\sR$ sector) as the boundary condition on the 10d gravitino around $S^{1}_{R}$ -- is encoded in the spectrum of conformal dimensions of the internal theory, and more crucially, it is sensitive to whether or not we fermionize the worldsheet theory using the $\Arf$ theory. We refer the reader to \cite{Atick:1988si,Acharya:2020hsc,Rohm:1983aq,Kounnas:1989dk} for previous discussions of the dependence of spacetime spin structure on worldsheet choices.

\section{Tachyonic Deformation}\label{sec:tachyon}
The internal CFT of the non-supersymmetric heterotic string in $d=9$ spacetime dimensions is a theory with central charge $(c_L, c_R) = (17,\frac{3}{2})$. We will consider the particular example of a spacetime theory constructed using the internal theory\footnote{The left-moving part of \eqref{eq:t17-3-by-2} was constructed in Section \ref{sec:tensoringe8}, see \eqref{eq:Theory-CF2}.}
\eqa{
  &\sT_2 \otimes \wt{\psi} = \big[\cbos_{2R} \otimes (E_8)_2 \times \lambda\big]\orb \mathsf{diag}\big(\IZ_2^\hp \times (-1)^{F}\big) \otimes \wt{\psi} &&= \sT_2 \otimes \wt{\psi} ~, \label{eq:t17-3-by-2}
}
(with $\wt{\psi}$ denoting a right-moving Majorana-Weyl fermion) after a GSO projection. The resulting theory has a gravitational anomaly $\nu = -2(c_L-c_R) = -31$. This internal $\cN=(0,1)$ worldsheet SCFT contributes a spacetime tachyon essential for establishing our main claim. In this section, we present an algorithm to count tachyons in the above theory and study its deformation by a tachyon vertex operator.

\subsection{Counting Tachyons}
In this section, we slightly generalize a result of \cite{BoyleSmith:2023xkd} about counting spacetime tachyons (based on an analysis of the worldsheet partition function) to non-chiral internal SCFTs. We will use this result to count tachyons in the spacetime theory, which correspond to $\sN\sS$ sector states with $L_{0} - \ov{L}_{0} = \frac{1}{2}$ as we demonstrate below.

Suppose $D$ denotes the number of compact dimensions of the heterotic string, so $D = 10-d$. Then the internal SCFT has central charge $(26-d,\frac{3}{2}(10-d)) = (16+D, \frac{3D}{2})$. In Appendix \ref{app:factoringintcft}, we give the expressions for the $\sN\sS$ sector partition functions of any two-dimensional $\mathcal{N}=(0,1)$ SCFT, see \eqref{eq:Z (0,1) NSNS refined} and \eqref{eq:Z (0,1) NSR refined}. For $(c_L, c_R) = (16+D, \frac{3D}{2})$, \eqref{eq:Z (0,1) NSNS refined} and \eqref{eq:Z (0,1) NSR refined} reduce to:
\eqa{
  &\sZ_{\sN\sS,\sN\sS}^{\sT} &&= q^{-\left(\frac{2}{3}+\frac{D}{24}\right)} \ov{q}^{\,-D/16} \sum_{Δ',\ov{Δ}'}N_{Δ',\ov{Δ}'} q^{Δ'}\ov{q}^{\ov{Δ}'} ~,  \label{eq:ZT (0,1) NSNS}\\
  &\sZ_{\sN\sS,\sR}^{\sT} &&= q^{-\left(\frac{2}{3}+\frac{D}{24}\right)} \ov{q}^{\,-D/16} e^{-4\pi \sfi /3}e^{\sfi \pi D/24} \sum_{Δ',\ov{Δ}'}N_{Δ',\ov{Δ}'}e^{2\pi \sfi  (Δ'-\ov{Δ}')} q^{Δ'} \ov{q}^{\ov{Δ}'} ~. \label{eq:ZT (0,1) NSR}
}
Consider a general spacetime theory $\cT$ given by a tensor product of two $\mathcal{N}=(0,1)$ SCFTs:
\eqa{
  &\cT &&: = \underbrace{(8-D) \cdot (X_L, X_R, \wt{\psi})}_{\substack{(c_L,c_R)=(8-D,\frac{3}{2}(8-D)) \\ \text{ and } \mathcal{N}=(0,1)}} \qquad \otimes \underbrace{\sT}_{\substack{(c_L,c_R)=(16+D,\frac{3D}{2}) \\ \text{ and } \mathcal{N}=(0,1)}} ~. \label{eq:general theory calT}
}
The GSO-projected partition function in the $\sN\sS$ sector is
\eqa{
&\left.\sZ^{\cT}_{\sN\sS}\right|_{\mathsf{GSO}} &&= \frac{(4\pi^2\mathsf{Im\,}\tau)^{-(8-D)/2}}{|\eta(\tau)|^{2(8-D)}}\frac{1}{2}\left[\overline{\left(\frac{\vartheta_3(\tau)}{\eta(\tau)}\right)}^{(8-D)/2}\sZ^{\sT}_{\sN\sS,\sN\sS} + \overline{\mathsf{a}}_{D}\overline{\left(\frac{\vartheta_4(\tau)}{\eta(\tau)}\right)}^{(8-D)/2}\sZ^{\sT}_{\sN\sS,\sR} \right] ~, \label{eq:GSO NS general}
} 
where $\overline{\mathsf{a}}_{D}$ is a $D$-dependent constant relating $\overline{\left(\frac{\vartheta_3(τ)}{η(τ)}\right)}^{(8-D)/2}$ to $\overline{\left(\frac{\vartheta_4(τ)}{η(τ)}\right)}^{(8-D)/2}$ under a $\mathcal{T}$-transformation:
\eqa{
  &\left(\frac{\vartheta_3(τ)}{η(τ)}\right)^{(8-D)/2} &&\xmapsto{\tau \mapsto \tau+1} \underbrace{e^{i\pi D/24} e^{-\sfi\pi/3}}_{\mathsf{a}_{D}}\left(\frac{\vartheta_4(τ)}{η(τ)}\right)^{(8-D)/2} ~. \label{eq:aD}
}
Therefore,
\eqa{
    \overline{\mathsf{a}}_{D} &= e^{-\sfi\pi D/24}e^{\sfi\pi/3} ~. \label{eq:aD-formula}
}
Note that $\overline{\mathsf{a}}_{D}$ combines with the phase in $\sZ_{\sN\sS,\sR}^{\sT}$ to give $-1$.

We begin by noting that
\eqa{
  &\overline{\left(\frac{\vartheta_3(τ)}{η(τ)}\right)}^{(8-D)/2} &&= \overline{q}^{(D-8)/48}\big( 1 + (8-d)\overline{q}^{1/2} + \tfrac{(8-D)(7-d)}{2}\overline{q}  + \cdots \big) := \overline{q}^{(D-8)/48}\sum_{n=0}^{\infty}\mathsf{t}_{n} \overline{q}^{n/2} ~. \label{eq:theta3exp1-8minusd} \\
  &\overline{\left(\frac{\vartheta_4(τ)}{η(τ)}\right)}^{(8-D)/2} &&= \overline{q}^{(D-8)/48}\big( 1 - (8-D)\overline{q}^{1/2} + \tfrac{(8-D)(7-D)}{2}\overline{q} + \cdots \big) := \overline{q}^{(D-8)/48} \sum_{n=0}^{\infty}(-1)^{n}\mathsf{t}_{n} \overline{q}^{n/2} ~. \label{eq:theta4exp1-8minusd}\\
  &\frac{1}{\eta(\tau)^{8-D}} &&= q^{(D-8)/24}\sum_{n=0}^{\infty}\mathsf{p}(n,8-D) q^{n} ~. \label{eq:one-over-eta-power}
}
Here $\mathsf{p}(n,k)$ is the number of partitions of a positive integer $n$ into $k$ parts. Also, the $\{\mathsf{t}_n\}$'s are $D$-dependent.

Ignoring the center-of-mass contribution, we have
\eqa{
  &\left.\sZ^{\cT}_{\sN\sS}\right|_{\mathsf{GSO}} &&\sim q^{\,-1}\ov{q}^{\,-\frac{1}{2}}\sum_{n,j,k=0}^{\infty}\sum_{Δ',\ov{Δ}'}\mathsf{t}_{n} \mathsf{p}(j,8-D)\mathsf{p}(k,8-D)N_{Δ',\ov{Δ}'}\tfrac{\big(1-e^{2\pi \sfi (Δ-\ov{Δ}')}(-1)^{n}\big)}{2}q^{Δ' + j}\,\ov{q}^{\ov{Δ}'+\frac{n}{2} + k} ~. \label{eq:Z NS final}
}
A state of $\sT$ with $(L_0, \ov{L}_0)$ eigenvalues $(Δ', \ov{Δ}')$ leads to a spacetime tachyon if
\eqa{
  Δ' + j - 1 &= \ov{Δ}' + \frac{n}{2} + k - \frac{1}{2} < 0 ~,
}
for some non-negative integers $n, j, k$. For sufficiently positive $n$, $j$, or $k$, this would imply a negative $Δ'$ or negative $\ov{Δ}'$, which are both impossible as the theory is unitary. Therefore, we must have $n = j = k = 0$, which implies that,
\beqa{
  0 &\leq Δ' < 1 ~, \quad 0 \leq \ov{Δ}' < \frac{1}{2} ~, \quad Δ' - \ov{Δ}' = \frac{1}{2} ~. \label{eq:tachyon condition}
}
This generalizes a result in \cite[Sec. 6]{BoyleSmith:2023xkd} to non-chiral CFTs. For $n = j = k = 0$, \eqref{eq:Z NS final} reduces to:
\eqa{
& \frac{1}{q \ov{q}^{1/2}}\sum_{Δ',\ov{Δ}'}N_{Δ',\ov{Δ}'}\left(\frac{1-e^{2\pi \sfi (Δ-\ov{Δ}')}}{2}\right)q^{Δ'}\ov{q}^{\ov{Δ}'} = \frac{1}{q \ov{q}}\sum_{Δ'}N_{Δ',Δ'-\frac{1}{2}}q^{Δ'}\ov{q}^{Δ'} ~. \label{eq:isolate Z NS final}
}
Therefore, a state with $(L_0, \ov{L}_0)$ eigenvalue $(Δ', Δ'-\frac{1}{2})$ contributes to tachyons if the GSO-projected partition function in the $\sN\sS$ sector can be reduced to the form \eqref{eq:isolate Z NS final} for $n = j = k = 0$, and then the number of tachyons is simply the value of the integer coefficient $N_{Δ',Δ'-\frac{1}{2}}$, which can be read off \eqref{eq:ZT (0,1) NSNS}. To illustrate this, we present two concrete examples.

\paragraph{The $D = 0$ case:}
For $D=0$, the theory $\sT$ of \eqref{eq:general theory calT} has central charge $(c_L, c_R)=(16,0)$, and one particular version of it is the $(E_8)_{2} \times \lambda$ theory\footnote{This is the $\overline{(E_8)_{2}} \times \lambda$ theory in the notation of \cite{BoyleSmith:2023xkd}.} for which, the $(\sN\sS,\sN\sS)$ partition function \eqref{eq:e8-level2-psi-NSNS} is
\eqa{
  &\sZ^{(E_8)_{2}\times \lambda}_{\sN\sS,\sN\sS} 
  &&= \big(\chi_{0}^{\Vir} + \chi_{1/2}^{\Vir}\big)\big(\chi_{\bm{1}}^{E_8} + \chi_{\bm{3875}}^{E_8}\big) ~. 
}
Its $q$-series expansion is
\eqa{
  &\sZ^{(E_8)_{2}\times \lambda}_{\sN\sS,\sN\sS} &&= q^{-2/3}\left(1 + q^{1/2} + 248 q + 4124 q^{3/2} + \cdots \right) ~. \label{eq:E8level2-ising-ZNSNS}
}
In this case, $\ov{Δ}' = 0$ (there are no right movers in $\sT$) and so $Δ' = \frac{1}{2}$ is the only contributor to tachyons.
Comparing \eqref{eq:E8level2-ising-ZNSNS} with \eqref{eq:ZT (0,1) NSNS}, we read off $N_{\frac{1}{2},0} = 1$. So there is a single spacetime tachyon. 

\paragraph{The $D = 1$ case:} For $D=1$, the theory $\sT$ of \eqref{eq:general theory calT} has central charge $(c_L, c_R)=(17,\frac{3}{2})$ and one realization of it is the tensor product theory
\eqa{
  &\sT_2 \otimes \wt{\psi} &&= \big[\cbos_{2R} \otimes (E_8)_2 \times \lambda\big]\orb \mathsf{diag}\big(\IZ_2^\hp \times (-1)^{F}\big) \otimes \wt{\psi} ~, \label{eq:Dequals1example}
}
where $\sT_2$ was discussed in Section \ref{sec:tensoringe8} (see \eqref{eq:Theory-CF2}) and $\wt{\psi}$ denotes a right-moving Majorana-Weyl fermion. We will show that this theory has a single tachyon for any odd integer value of the momentum modenumber, provided the radius is large enough. Readers who wish to avoid the details of this analysis can safely refer to the final result \eqref{eq:radiusresult} and skip to the next subsection.

The partition functions of $\sT_2$ were derived in \eqref{eq:Z-F2-NSNS}--\eqref{eq:Z-F2-RR}. For a right-moving Majorana-Weyl fermion,
\eqa{
  \sZ_{\sN\sS,\sN\sS}^{\wt{\psi}} &= \overline{\chi}_{0}^{\Vir} + \overline{\chi}_{1/2}^{\Vir} ~,\\
  \sZ_{\sN\sS,\sR}^{\wt{\psi}} &= e^{\sfi \pi/24}\big(\overline{\chi}_{0}^{\Vir} - \overline{\chi}_{1/2}^{\Vir}\big) ~,
}
where the phase factor in the second equation follows from $\sfa_{D=7}$, see \eqref{eq:aD-formula}. Therefore, for the product theory $\sT := \sT_2 \otimes \wt{\psi}$,
\eqa{
  &\sZ_{\sN\sS,\sN\sS}^{\sT} &&= \left(\frac{1}{2}\sZ^{\mathsf{cb}}_{\mathsf{0},\mathsf{0}}\left(\chi_{0}^{\Vir}\chi_{\bm{1}}^{E_8} + \chi_{1/2}^{\Vir}\chi_{\bm{1}}^{E_8} + \chi_{0}^{\Vir}\chi_{\bm{3875}}^{E_8} + \chi_{1/2}^{\Vir}\chi_{\bm{3875}}^{E_8} \right)\right. \nonumber\\
  & &&\quad + \left.\frac{1}{2}\sZ^{\mathsf{cb}}_{\mathsf{0},\mathsf{1}}\left(\chi_{0}^{\Vir}\chi_{\bm{1}}^{E_8} - \chi_{1/2}^{\Vir}\chi_{\bm{1}}^{E_8} - \chi_{0}^{\Vir}\chi_{\bm{3875}}^{E_8} + \chi_{1/2}^{\Vir}\chi_{\bm{3875}}^{E_8} \right) + \sZ_{\mathsf{1},\mathsf{0}}^{\mathsf{cb}}\chi_{1/16}^{\Vir}\chi_{\bm{248}}^{E_8}\right)\left(\overline{\chi}_{0}^{\Vir} + \overline{\chi}_{1/2}^{\Vir}\right) ~, \label{eq:ZT-pf-1}
}
which is a function of $(q, \ov{q})$ and $R$ (the radius of the compact boson is $2R$). The $q$-series of the $(E_8)_2$ and Ising characters (see Appendix \ref{app:compchar}) yield
\eqa{
  & \chi_{0}^{\Vir}\chi_{\bm{1}}^{E_8} + \chi_{1/2}^{\Vir}\chi_{\bm{1}}^{E_8} + \chi_{0}^{\Vir}\chi_{\bm{3875}}^{E_8} + \chi_{1/2}^{\Vir}\chi_{\bm{3875}}^{E_8} &&= q^{-2/3}\left(1 + q^{1/2} + 248\,q + 4124\,q^{3/2} + \cdots\right) \nonumber\\
  & &&:= q^{-2/3}\sum_{n=0}^{\infty}\mathsf{r}_{n}q^{n/2} ~,\\
  & \chi_{0}^{\Vir}\chi_{\bm{1}}^{E_8} - \chi_{1/2}^{\Vir}\chi_{\bm{1}}^{E_8} - \chi_{0}^{\Vir}\chi_{\bm{3875}}^{E_8} + \chi_{1/2}^{\Vir}\chi_{\bm{3875}}^{E_8} &&= q^{-2/3}\left(1 -q^{1/2} + 248\,q - 4124\,q^{3/2} + \cdots\right) \nonumber\\
  & && := q^{-2/3}\sum_{n=0}^{\infty}\mathsf{r}_{n}(-1)^{n}q^{n/2} ~,\\
  & \chi_{1/16}^{\Vir}\chi_{\bm{248}}^{E_8} &&= q^{\,+1/3}\left( 248 + 34752 q + 1057504 q^2 + \cdots \right)\nonumber\\
  & && := q^{+1/3}\sum_{n=0}^{\infty}\mathsf{s}_{n}q^{n} ~.
 }
 The integer coefficients $\mathsf{r}_{n}$ and $\mathsf{s}_{n}$ can be read off from the corresponding series expansions. From \eqref{eq:theta3exp1-8minusd},
 \eqa{
  & \overline{\chi}_{0}^{\Vir} + \overline{\chi}_{1/2}^{\Vir} &&= \overline{\left(\frac{\vartheta_3(\tau)}{\eta(\tau)}\right)}^{1/2} = \ov{q}^{\,-1/48}\left(1 + \ov{q}^{1/2} + \ov{q}^{3/2} + \ov{q}^2 + \cdots\right) = 
  \ov{q}^{\,-1/48}\sum_{n=0}\mathsf{t}_{n}\ov{q}^{n/2} ~.
 }
 Using \eqref{eq:one-over-eta-power}, 
 \eqa{
  &\frac{1}{|\eta(\tau)|^2} &&= q^{-1/24}\ov{q}^{-1/24}\sum_{n',m'=0}^{\infty}\mathsf{p}(n',1)\mathsf{p}(m',1) q^{n'} \ov{q}^{m'} ~.
}
Therefore, we can rewrite \eqref{eq:z-cbosR-ab} for $\cbos_{2R}$ as:
\eqa{
  & \sZ_{\mathsf{a,b}}(\cbos_{2R}; \IZ_{2}^{\hp}) &&= \sum_{n',m' = 0}^{\infty}\mathsf{p}(n',1)\mathsf{p}(m',1)\sum_{\substack{m \in \bZ \\ n \in \bZ + \frac{1}{2}\mathsf{a}}}(-1)^{m \mathsf{b}}q^{\frac{1}{4}\left(\frac{m}{2R} + 2nR\right)^2-\frac{1}{24}}\ov{q}^{\frac{1}{4}\left(\frac{m}{2R} - 2nR\right)^2-\frac{1}{24}} ~. \label{eq:z-cbosR-ab-repeated}
}
Finally, \eqref{eq:ZT-pf-1} can be written as
\eqa{
 & \bm{\sZ}_{\sN\sS,\sN\sS}^{\sT} &&= q^{-\frac{17}{24}}\ov{q}^{\,-\frac{1}{16}}\bigg(\sum_{n,m\in\mathbb{Z}}\sum_{\substack{ n',m',\\\ell',\kappa'=0}}^{\infty} \mathsf{p}(n',1)\mathsf{p}(m',1)\mathsf{r}_{\ell'}\mathsf{t}_{\kappa'}\bigg(\tfrac{1 + (-1)^{m + \ell'}}{2}\bigg) q^{\frac{1}{4}\left(\frac{m}{2R} + 2nR\right)^2 + m' + \frac{1}{2}\ell'}\ov{q}^{\frac{1}{4}\left(\frac{m}{2R} - 2nR\right)^2 + n' + \frac{1}{2}\kappa'} \nonumber\\
 & &&\quad + \sum_{n,m\in\mathbb{Z}}\sum_{\substack{ n',m',\\\ell',\kappa'=0}}^{\infty}\mathsf{p}(n',1)\mathsf{p}(m',1)\mathsf{s}_{\ell'}\mathsf{t}_{\kappa'}q^{\frac{1}{4}\left(\frac{m}{2R} + 2\left(n+\frac{1}{2}\right)R\right)^2 + m'+ \frac{1}{2}\ell' + 1 }\ov{q}^{\frac{1}{4}\left(\frac{m}{2R} - 2\left(n+\frac{1}{2}\right)R\right)^2 + n' + \frac{1}{2}\kappa'}\bigg) ~. \label{eq:3p32}
}
There are two classes of states that contribute to the partition function:
\begin{itemize}
    \item Class $\sI$: states with $(Δ', \ov{Δ}') = \left(\frac{1}{4}\left(\frac{m}{2R} + 2nR\right)^2 + m' + \frac{1}{2}\ell' ~\bm{,}~ \frac{1}{4}\left(\frac{m}{2R}-2nR\right)^2 + n' + \frac{1}{2}\kappa'\right)$ ~,
    \item Class $\sI\sI$: states with $(Δ', \ov{Δ}') = \left(\frac{1}{4}\left(\frac{m}{2R} + (2n+1)R\right)^2 + m' + \frac{1}{2}\ell' + 1 ~\bm{,}~ \frac{1}{4}\left(\frac{m}{2R}-(2n+1)R\right)^2 + n' + \frac{1}{2}\kappa'\right)$.
\end{itemize}
It is easy to check that $Δ'-\ov{Δ}'$ is valued in $\frac{1}{2}\mathbb{Z}$ for both classes.
For tachyons to exist, the condition \eqref{eq:tachyon condition} must be satisfied. This translates to the following conditions:
\begin{enumerate}\itemsep -5pt
    \item Class $\mathsf{I}$: the following conditions must be concurrently satisfied:
    \begin{itemize}\itemsep -1pt
      \item[$(\mathsf{I.A})$] $mn + m' - n' + \frac{1}{2}\ell' - \frac{1}{2}\kappa' = \frac{1}{2}$ 
      \item[$(\mathsf{I.B})$] $0 \leq \frac{1}{4}\left(\frac{m}{2R}+2nR\right)^2 + m' + \frac{1}{2}\ell' < 1$
      \item[$(\mathsf{I.C})$] $0 \leq \frac{1}{4}\left(\frac{m}{2R}-2nR\right)^2 + n' + \frac{1}{2}\kappa' < \frac{1}{2}$ 
    \end{itemize}
    \item Class $\mathsf{II}$: the following conditions must be concurrently satisfied:
    \begin{itemize}
      \item[$(\mathsf{II.A})$] $mn + m' - n'+ \frac{1}{2}m + \frac{1}{2}\ell' - \frac{1}{2}\kappa' = -\frac{1}{2}$ 
      \item[$(\mathsf{II.B})$] $0 \leq \frac{1}{4}\left(\frac{m}{2R}+(2n+1)R\right)^2 + m' + \frac{1}{2}\ell' + 1 < 1$ 
      \item[$(\mathsf{II.C})$] $0 \leq \frac{1}{4}\left(\frac{m}{2R}-(2n+1)R\right)^2 + n' + \frac{1}{2}\kappa' < \frac{1}{2}$ 
    \end{itemize}
  \end{enumerate}
  States from Class $\sI\sI$ do not contribute tachyons because condition $(\mathsf{II.B})$ is always violated. Therefore, we restrict our attention to Class $\sI$. We further observe that:
\begin{enumerate}\itemsep 0pt
  \item For a physical state $(m+\ell')$ must be even, due to the projector in \eqref{eq:3p32}. So $m$ and $\ell'$ are both even or both odd.
  \item If $\ell'$ and $\kappa'$ both vanish, condition $(\mathsf{I.A})$ cannot hold. 
  \item If $\kappa' \geq 1$, condition $(\mathsf{I.C})$ cannot hold. Therefore we must have $\underline{\kappa' = 0}$. Then, by the previous observation, $\ell' \neq 0$. 
  \item With $\kappa' = 0$, condition $(\mathsf{I.C})$ cannot hold if $n' \geq 1$. Therefore we must have $\underline{n' = 0}$.
  \item Condition $(\mathsf{I.B})$ cannot be satisfied if $\ell' \geq 2$. And since point 3. required $\ell' \neq 0$, we must have $\underline{\ell' = 1}$. Then point 1. implies that $\underline{m \text{ must be odd}}$.
  \item With $\ell'=1$, condition $(\mathsf{I.B})$ implies that $\frac{1}{4}\left(\frac{m}{2R}+2nR\right)^2 + m' < \frac{1}{2}$. This cannot hold if $m' \geq 1$. Therefore we must have $\underline{m' = 0}$.
  \item Finally, with $m' = n' = \kappa'=0$ and $\ell'=1$, condition $(\mathsf{I.A})$ reduces to $\underline{mn = 0}$, which implies that at least one of $m$ or $n$ is zero. Since point 5. required that $m$ be odd, we take $m \neq 0$. Therefore $\underline{n = 0}$.
\end{enumerate}
To summarize,
\eqa{
    \kappa' &= 0 ~,~ n' = 0 ~,~ \ell' = 1 ~,~ m' = 0 ~,~ m \text{ is odd} ~,~ \text{ and }  mn = 0 ~.
}
The number of tachyons with $(n, m, n', m', \ell', \kappa') = (0, m, 0, 0, 1, 0)$ is 
\eqa{
    \mathsf{p}(n',1)\mathsf{p}(m',1)\mathsf{r}_{\ell'}\mathsf{t}_{\kappa'}\left(\frac{1 + (-1)^{m+\ell'}}{2}\right) &= \mathsf{p}(0,1)\mathsf{p}(0,1)\mathsf{r}_1 \mathsf{t}_0 = 1 ~.
}
Therefore, there is single tachyon for any $m \in 2\mathbb{Z}+1$ provided the radius satisfies
\beqa{
    R &> \frac{|m|}{2\sqrt{2}} ~, \quad \text{$m$ odd} ~. \label{eq:radiusresult}
}

\subsection{Deformation by the Tachyon Vertex Operator and RG Flow}\label{sec:tachyondeformationoperator}
We continue our discussion of the $D=1$ worldsheet theory $\sT_2 \otimes \wt{\psi}$ theory defined in \eqref{eq:Dequals1example}. As we saw, the spacetime spectrum contains a tachyon. By the state-operator map, it corresponds to a tachyon vertex operator in the worldsheet SCFT. We will deform the worldsheet theory by this operator, which is characterized by the following properties: (1) It is a Virasoro primary with conformal weight $\left(\frac{m^2}{16R^2} + \frac{1}{2}, \frac{m^2}{16R^2}\right)$ where $m$ is odd and\footnote{Note that the conformal weights are upper-bounded.} $R > \frac{|m|}{2\sqrt{2}}$, and (2) it is a superVirasoro primary.

The worldsheet superfields in $\cN=(0,1)$ superspace parametrized by $(z, \ov{z}, \ov{\theta})$ are
\eqa{
    \bm{X}^{\mu}(z, \ov{z}, \ov{\theta}) &= X^{\mu}(z, \ov{z}) + \sfi\ov{\theta}\,\wt{\psi}^{\mu}(\ov{z}) ~, \quad (\text{scalar multiplet})\\
    \bm{\lambda}^{A'}(z, \ov{\theta}) &= \lambda^{A'}(z) + \ov{\theta}\,F^{A'}(z) ~, \quad (\text{Fermi multiplet})
}
where $F^{A'}$ is an auxiliary field. Here $A' = (1, \ldots, 32)$ and $A = (1, \ldots, 31)$, and we will isolate the $A' = 32$ lone Fermi multiplet, conventionally written without a superscript.  Next, we add a deformation\footnote{As a superfield, $\bm{\mathcal{O}}(z, \ov{z}) = \mathcal{O}_{0}(z,\ov{z}) + \ov{\theta}\,\mathcal{O}_{1}(z, \ov{z})$, where the components satisfy $L_{0}\cdot\cO_{0} = h\cO_{0}$ , $L_{0}\cdot\cO_{1} = (h+\frac{1}{2})\cO_{1}$ , $L_{m}\cdot \cO_{0,1} = 0$ for $m \geq 1$ (idem for the right-handed components, with $L \leftrightarrow \wt{L}$ and $h \leftrightarrow \wt{h} := h-\frac{1}{2}$), $\wt{G}_{-1/2} \cO_{0} = \cO_{1}$, $\wt{G}_{r}\cdot \cO_{0} = 0$ for $r \geq \frac{1}{2}$ , $\wt{G}_{-1/2}\cdot \cO_{1} = \ov{\partial}\cO_{0}$ , $\wt{G}_{1/2}\cdot\cO_{1} = 2\wt{h}\cO_{0}$, and $\wt{G}_{r}\cdot\cO_{1} = 0$ for $r \geq \frac{3}{2}$. We follow the conventions of \cite{Polchinski:1998rr}.}
\eqa{
   \Delta S &:= \mu\int d^2 z\, d\ov{\theta}\,\bm{\mathcal{O}}(z, \ov{z}, \ov{\theta}) ~,
}
to the worldsheet action of the heterotic string.\footnote{See \cite{Hull:1985jv} and \cite[Sec.12.3]{Polchinski:1998rr} for a discussion of the $\cN=(0,1)$ sigma model action.} 
In general, turning on an $\cN=(0,1)$ superpotential,
\eqa{
     \bm{\mathcal{O}}(z,\ov{z},\ov{\theta}) := \bm{\lambda}\,J(\bm{X}) ~,\label{eq:superpot}
}
where $\bm{\lambda}$ is the Fermi superfield corresponding to the lone Majorana-Weyl fermion on the worldsheet, and $J$ is a real function of the superfield $\bm{X}$, yields
\eqa{
    \Delta S &= \mu \int d^{2}z \,d\ov{\theta}\,\bm{\lambda}\,J(\bm{X}) = 
    \mu \int d^{2}z\left( F J(X) - \sfi \lambda\widetilde{\psi}^{\mu}\frac{\partial J}{\partial X^{\mu}}\right) ~.\label{eq:4p15}
}
Integrating out the auxiliary field $F$ using\footnote{Here, we are setting the pullbacks of the connections for both $\lambda^{A}$ and $\lambda$ to zero. It may be interesting to study generalizations where the corresponding gauge fields are turned on.} $F = \frac{1}{2}\mu J(X)$, this gives a scalar potential $V(X) = \frac{1}{2}\mu^2 J^2(X)$. For the tachyon vertex operator deformation,  we take $J(\bm{X}) = e^{\sfi\bm{k}\cdot\bm{X}}$, for which $\Delta S = \mu\int d^{2}z\, \big(F + \lambda k_{\mu}\wt{\psi}^{\mu}\big)e^{\sfi k\cdot X }$. The integrand is the (off-shell) tachyon vertex operator for the non-supersymmetric heterotic string \cite{Horava:2007yh,Horava:2007hg,Hellerman:2007zz}. We will add the real part of this to the worldsheet, which reads
\eqa{
    \Delta S &= \mu\int d^{2}z\, \big(F \cos(k\cdot X) + \sfi\lambda k_{\mu}\wt{\psi}^{\mu}\sin(k\cdot X)\big) ~. \label{eq:off-shell-delta-S}
}
This corresponds to $J(X) = \cos(k\cdot X)$, and integrating out the auxiliary field gives
\eqa{
    \Delta S &= \mu \int d^{2}z\left[ \frac{\mu}{2}\cos^2(k\cdot X) + \sfi\lambda k_{\mu}\wt{\psi}^{\mu}\sin(k\cdot X)\right] ~. \label{eq:cont-def-DeltaS}
}
Let us consider a single compactified $X$ coordinate, a.k.a. a compact boson (with its superpartner $\wt{\psi}$ being a right-moving Majorana-Weyl fermion). This is a relevant deformation if\footnote{In our conventions, $\alpha'=1$, so the operator $e^{i\bm{k}\cdot\bm{X}}$ has conformal dimension $(\frac{\bm{k}^2}{4}, \frac{\bm{k}^2}{4})$.}
\eqa{
   \frac{k^2}{4} &= \frac{m^2}{16 R^2} < \frac{1}{2} ~,
}
that is, if,
\eqa{
 k &= \pm \frac{|m|}{4 R} \in (-\sqrt{2}, \sqrt{2}) ~.
}
Without loss of generality, we can pick the positive sign. Therefore, the scalar potential for $X$ is
\eqa{
  V(X) &= \frac{1}{2}\mu^2\cos^2\left(\frac{|m|X}{4R}\right)  \quad (m \in 2\IZ + 1) ~. \label{eq:4p28}
}
The minima of $V(X)$ are $X = \frac{2\pi R}{|m|}(2 p + 1)$ where $p \in \mathbb{Z}$. As $X$ is compact, there is an odd number $|m|$ of vacua.\footnote{Strictly speaking, there is no spontaneous symmetry breaking at finite volume due to tunneling. We assume that $R$ is large enough to justify the large-volume argument to claim that there are $|m|$ vacua.} This deformation breaks the $U(1)$ rotation symmetry to a discrete $\mathbb{Z}_{|m|}$ symmetry, which simply permutes the $|m|$ vacua. It is clear (for example, by expanding about one of the $|m|$ vacua) that the scalar $X$, its superpartner $\wt{\psi}$, and the lone fermion $\lambda$ all acquire a mass, with a scale set by $\mu$. Below this mass scale, the light degrees of freedom correspond to the $(E_8)_2$ current algebra fermions. 

Therefore, adding this relevant deformation to the $\sT_2 \otimes \wt{\psi}$ theory triggers an RG flow to the $(E_8)_2$ theory. Combining this with the previously established T-duality of $\sT_1 \otimes \wt{\psi}$ and $\sT_2 \otimes \wt{\psi}$, we have effectively shown that the theory $\sT_1 \otimes \wt{\psi}$ is continuously connected to the $(E_8)_2$ theory.

\section{Implication for Topological Modular Forms}\label{sec:tmf}
As remarked in the introduction, there is a conjectured relation due to the work of Segal \cite{Segal88,Segal2007}, and Stolz and Teichner \cite{StolzTeichner1,StolzTeichner2}, between 2d $\cN=(0,1)$ SQFTs and the generalized cohomology theory known as Topological Modular Forms (TMF) \cite{Hopkins95,Hopkins2002,Hopkins:2002rd,Goerss:2009,Lurie2009,Douglas2014-cl,BrunerRognes,Gukov:2018iiq,Tachikawa:2021mvw,Tachikawa:2023lwf,Tachikawa:2023nne}. We begin with a very rudimentary review to motivate a consequence of the above results for TMF. If $\cS_{-\nu}$ denotes the space of all 2d $\cN = (0,1)$ SQFTs with an anomaly $\nu \in \IZ$ and $\sX$ is a topological space, the Segal-Stolz-Teichner conjecture states that\footnote{There exist equivariant and twisted versions, see \cite{Gukov:2018iiq,Tachikawa:2021mby}.}
\begin{align}
 \hspace{-0.1in}\mathsf{TMF}^{-\nu}(\sX) &\cong \Bigg\{\begin{array}{c} \text{\small families of}\\ \text{\small 2d $\cN=(0,1)$ SQFTs $\in S_{-\nu}$} \\ \text{\small parametrized by $\sX$} \end{array}\Bigg\}\left/\Bigg\{\begin{array}{c}\text{\small continuous}\\ \text{\small SUSY-preserving}\\ \text{\small deformations}\end{array}\Bigg\}\right. = [\sX, \cS_{-\nu}] 
 :=\mathsf{Map}(\sX, \cS_{-\nu})/\sim. \label{eq:stolz-teichner}
\end{align}
 Physically, this means that the deformation class of a theory in $S_{-\nu}$ parametrized by $\sX$ is a representative of a cocycle for $\mathsf{TMF}^{-\nu}(\sX)$. More formally, \eqref{eq:stolz-teichner} states that deformation classes of 2d $\cN=(0,1)$ SQFTs with anomaly coefficient $\nu \in (I_{\IZ}\Omega^{\mathsf{spin}})^{4}(\mathrm{pt}) \cong \IZ$ (where $I_{\IZ}\Omega^{\mathsf{spin}}$ denotes the Anderson dual\footnote{The appearance of the Anderson dual can be traced back to the work of Freed and Hopkins \cite[Sec. 9]{Freed:2016rqq}, who described a natural transformation $\alpha: \mathsf{KO}^{d-2}(\sX) \to (I_{\IZ}\Omega^{\mathsf{spin}})^{d+2}(\sX)$ that maps a $d$-dimensional massless fermionic theory to the deformation class of its invertible anomaly field theory.} of the spin-bordism group) form the Abelian group $\mathsf{TMF}^{-\nu}(\sX)$. When $\sX = \mathrm{pt}$, this reduces to an isomorphism,
 \eqa{
  \mathsf{TMF}^{\bullet} := \mathsf{TMF}^{\bullet}(\mathrm{pt}) &\cong \pi_{0}(\cS_{\bullet}) ~, \label{eq:SST-Isomorphism}
 }
 between TMF classes and the set of path-connected components of $\cS_{\bullet}$. $\mathsf{TMF}^{\bullet}$ is an $E_{\infty}$-ring spectrum, graded by a degree $-\nu \in \IZ$. Therefore,
 \eqa{
  \pi_{\nu}(\mathsf{TMF}) = \mathsf{TMF}_{\nu} = \mathsf{TMF}^{-\nu} = \pi_{0}(\cS_{-\nu}) ~. \label{eq:st-point}
 }
 The ring structure on $\cS_{\bullet}$ is given by defining the sum of theories $\sT_\alpha + \sT_\beta$ (with the Hilbert space being the direct sum $\cH(\sT_\alpha) \oplus \cH(\sT_\beta)$), the product of theories $\sT_\alpha \otimes \sT_\beta$ (noninteracting stacking, with the Hilbert space being the tensor product $\cH(\sT_\alpha) \otimes \cH(\sT_\beta)$), and a parity-reversal operation `$-$' (under which the fermionic parity of the $\sR$-sector vacuum is flipped; note that $-\sT_\alpha = \sT_\alpha \otimes \Arf$ is simply a product with the $\Arf$ theory). The first two operations correspond, respectively, to the sum and product operations in $\mathsf{TMF}^{\bullet}$. On $\cS_{\bullet}$, the multiplicative unit $\bm{1}$ is a theory with a single gapped vacuum and zero degrees of freedom, such that $\bm{1} \otimes \sT_\alpha \cong \sT_\alpha \otimes \bm{1} \cong \sT_\alpha$, whereas the additive unit $\bm{0}$ is the empty theory with no vacuum and zero degrees of freedom, such that $\bm{0}\otimes \sT_\alpha \cong \sT_\alpha \otimes \bm{0} \cong \bm{0}$. With this structure, it is clear that the set $\pi_{0}(\cS_{\bullet})$ is, in fact, an Abelian group, and \eqref{eq:st-point} is therefore an isomorphism of Abelian groups.

 In a series of works \cite{Tachikawa:2021mvw,Tachikawa:2021mby,Tachikawa:2023lwf,Tachikawa:2023nne}, this conjectured relationship between 2d $\cN=(0,1)$ theories and TMF has been extensively fleshed out from physical and mathematical viewpoints, in the context of general 2d $\cN=(0,1)$ SQFTs, and also those that appear as internal theories for the $d$-dimensional heterotic string. In the latter setting, the internal 2d $\cN=(0,1)$ theory has anomaly coefficient $\nu = -2(c_L-c_R) = -22-d$. Let us briefly summarize some salient points of \cite{Tachikawa:2023lwf}. For any degree $\nu \in \IZ$, there is a homomorphism between homotopy groups,\footnote{The homomorphism \eqref{eq:TMFphimap} is induced by the morphism at the level of spectra, $\mathsf{TMF} \xrightarrow{\sigma} \mathsf{KO}((q))$, described in \cite[App. A]{HillLawson}. Here, $\mathsf{KO}((q))$ denotes a Laurent series with coefficients valued in $\mathsf{KO}$. As explained in \cite{Tachikawa:2021mby}, the morphism $\sigma$ amounts to putting a 2d SQFT on $S^1$ so that it is equivariant with respect to the $U(1)$ action, and the power of $q$ then specifies the weight of this action. The morphism \eqref{eq:TMFphimap} physically corresponds to considering the right-moving ground states of the internal SCFT in the $\mathsf{R}$ sector, from which one can infer the massless spacetime fermionic spectra \cite{Tachikawa:2021mby,Tachikawa:2023nne}.}
 \eqa{
    \Phi : \pi_{\nu}\mathsf{TMF} \longrightarrow \pi_{\nu}\mathsf{KO}((q)) ~, \label{eq:TMFphimap}
 }
 with kernel and image denoted respectively by $A_{\nu}$ and $U_{\nu}$. A given 2d $\cN=(0,1)$ SQFT $\cT$ of degree $\nu$ specifies, by the Segal-Stolz-Teichner isomorphism \eqref{eq:SST-Isomorphism}, a class $[\cT] \in \mathsf{TMF}_{\nu}$, and one can compute its image $\Phi([\cT])$ under this map. When $\nu \equiv 0, 4 \mathsf{\,\,mod\,\,}8$, for which $\mathsf{KO}_{\nu} \cong \IZ$, the image is the $q$-expansion of the standard elliptic genus of $\cT$ \cite{Witten:1986bf}. When $\nu \equiv 1, 2 \mathsf{\,\,mod\,\,}8$, for which $\mathsf{KO}_{\nu} \cong \IZ/2\IZ$ (which we will always denote by $\IZ_2$ below), the image is the $q$-expansion of the mod-2 elliptic genus of $\cT$, which was studied in \cite{Tachikawa:2023nne}. If the image $\Phi([\cT])$ is nonzero in $U_{\nu}$, the theory has a nonzero mod-2 elliptic genus, whereas if the image vanishes (which means that the corresponding $\mathsf{TMF}$ class lives in $A_{\nu}$), the theory has a zero mod-2 elliptic genus.

 For us, the pertinent case is $d = 9$, when $\nu = -31 \equiv 1 \mathsf{\,\,mod\,\,}8$. Indeed, the T-dual pairs $\sT_1\otimes \wt{\psi}$ and $\sT_2\otimes \wt{\psi}$ as well as the $(E_8)_2$ theory, all belong to the space $\cS_{-31}$. First of all, $\mathsf{TMF}_{-31}$ splits as the direct sum $\mathsf{TMF}_{-31} = A_{-31} \oplus U_{-31}$ (see \cite[Thm. 9.26]{BrunerRognes} and \cite[Prop. E.11]{Tachikawa:2023lwf}). The TMF class defined by,
 \eqa{
  \hspace{-0.1in}x_{-31} &:= \left[\left(\begin{array}{c}\text{\small the model obtained by fibering $(E_8)_1 \times (E_8)_1$ over the $\cN=(0,1)$} \\ \text{\small $S^1$ sigma model with antiperiodic spin structure such that} \\ \text{\small the two $E_8$ factors are exchanged upon going around $S^1$} \end{array}\right)\right] = \big[\,\sT_1 \otimes \wt{\psi}\,\big] \in \mathsf{TMF}_{-31} ~,
 }
 has been shown by Tachikawa and Yamashita in \cite[Sec. 4.4]{Tachikawa:2023lwf} to be the generator of $A_{-31} \cong \IZ_2$. The proof relies on interpreting $x_{-31}$ in terms of a certain power operation for the Adams spectral sequence for $\mathsf{TMF}$. In other words, our theory $\sT_1 \otimes \wt{\psi}$ is the generator $x_{-31}$ of $A_{-31}$. In \cite{Tachikawa:2023nne}, Tachikawa, Yamashita, and Yonekura further show that $U_{-31}$ -- which is pure torsion by definition -- is, in particular, the mod-$2$ reduction of an integral modular form of weight $-16$. To summarize, $\mathsf{TMF}_{-31}$ is pure torsion. 

 In general, suppose an ultraviolet (UV) 2d $\cN=(0,1)$ theory $\cT_{\sU\sV} \in \cS_{-\nu}$ has $N$ vacua labeled by $i = 1, \ldots, N$ in infinite volume. Suppose that on the $i^{th}$ vacuum, the theory in the infrared (IR) limit is $\cX_i$. Then the total IR theory is the sum of $\{\cX_i\}_{i=1}^{N}$ in $\cS_{-\nu}$. So, for the corresponding TMF classes, we have,
 \eqa{
  [\cT_{\sU\sV}] &= \sum_{i=1}^{N} [\cX_i] \quad \text{ in } \mathsf{TMF}_{\nu} ~. \label{eq:sumuvir}
 }
If we take as UV theory the $\sT_1 \otimes \wt{\psi}$ model with the tachyon vertex deformation of Section \ref{sec:tachyondeformationoperator} added, there are $|m|$ vacua where $m$ is odd.\footnote{Here we are implicitly invoking T-duality to replace $\sT_1 \otimes \wt{\psi}$ with $\sT_2 \otimes \wt{\psi}$.} We have shown that this theory is continuously connected to the $(E_8)_2$ theory and that in each vacuum, the theory in the IR limit is a copy of the $(E_8)_2$ theory.\footnote{\label{foot:moregeneral}In more general situations, for a given UV theory with a given degree $\nu$ -- the gravitational anomaly -- one might have distinct summands in \eqref{eq:sumuvir}, but each summand will have the same degree $\nu$ due to anomaly matching.} It follows that,
\eqa{
  x_{-31} = \big[\sT_1 \otimes \wt{\psi} + \text{tachyon deformation}\big] &= |m|\cdot [ (E_8)_2] ~,
}
where $[(E_8)_2]$ denotes the nontrivial TMF class of the $(E_8)_2$ theory in $\mathsf{TMF}_{-31}$. But since $m$ is always odd, 
\beqa{
  2\cdot [(E_8)_2] = 0 ~.
}
So, in effect, we have shown that 
\beqa{
 & x_{-31} &&= [ (E_8)_2] \in A_{-31} = \mathsf{ker}\big(\pi_{-31}\mathsf{TMF} \to \pi_{-31}\mathsf{KO}((q))\big) \cong \IZ_2 ~. \label{eq:x-31 is in A-31}
}
This result has a nice consistency check based on the mod-2 elliptic genus. For a left-moving spin CFT with degree $\nu \equiv 1 \mathsf{\,\,mod\,\,}8$, the mod-2 elliptic genus is simply the mod-2 reduction of the $q$-expansion of $\chi_{\sR,\sN\sS}$ \cite{Tachikawa:2023nne}. In our case, in the deep IR, the light degrees of freedom at any one of the odd number of vacua are those of the $(E_8)_2$ current algebra, which is a left-moving CFT with central charge $(\frac{31}{2},0)$ and degree $\nu = -31 \equiv 1 \mathsf{\,\,mod\,\,}8$, for which $\chi_{\sR,\sN\sS} = \chi_{\bm{248}}^{E_8}$. Using \eqref{eq:chi-e8level2-248}, the $q$-expansion of this is\footnote{The leading power of $q$ reflects the conformal weight $\frac{15}{16}$ of the integrable representation $\underline{\bm{248}}$ (adjoint) of $(E_8)_2$ and the central charge $\frac{31}{2}$.} 
\begin{align}
\hspace{-0.02in}\chi_{\bm{248}}^{E_8} &= q^{7/24}\left(248 + 34504\,q + 1022752\,q^2 + 16275496\,q^3 + 179862248\,q^4 + 1551303736\,q^5 + \cO(q^6) \right) ~.
\end{align}
The $q$-expansion has even coefficients\footnote{Furthermore, the coefficients are multiples of $8$. Both facts have been verified numerically to $\cO(q^{800})$.} and hence, the mod-2 elliptic genus vanishes. Thus the TMF class $[(E_8)_2]$ of the $(E_8)_2$ theory lives in the kernel $A_{-31}$, confirming \eqref{eq:x-31 is in A-31}.

This proves that the $(E_8)_2$ theory corresponds to the unique nonzero class in $\mathsf{TMF}_{-31}$ with zero mod-2 elliptic genus, thus giving a physical derivation of Conjecture A.10 of \cite{Tachikawa:2023lwf}.

\section{Conclusions and Future Directions\label{sec:conclusions}}
In this paper, we have established that the torsional class in $\mathsf{TMF}_{-31}$ corresponding to the $(E_8)_1 \times (E_8)_1$ theory (which was discussed in \cite{Tachikawa:2023lwf}) admits another distinguished representative, namely the $(E_8)_2$ theory. Our conclusion relies on the fact that these two theories are continuously connected in the space of 2d $\cN=(0,1)$ SQFTs with anomaly coefficient $\nu = -31$ (as shown in Section \ref{sec:tachyondeformationoperator}), and that Topological Modular Forms are complete deformation invariants of 2d $\cN=(0,1)$ SQFTs under the assumption of the Segal-Stolz-Teichner conjecture.\footnote{The term "complete deformation invariant" means the following: two 2d $\cN=(0,1)$ SQFTs (with the same anomaly coefficient $\nu$) are in the same deformation class or equivalently, are homotopic (and hence can be smoothly deformed into each other) if their images in $\mathsf{TMF}^{-\nu}$ under the Segal-Stolz-Teichner isomorphism \eqref{eq:defclass} coincide.} In particular, we have used physical arguments to give a homotopy between these theories \cite{Gaiotto:2019asa} defined in terms of a relevant tachyon vertex deformation. This work identifies the TMF class of the $(E_8)_2$ theory with that of the $(E_8)_1 \times (E_8)_1$ theory. This is not a mere curiosity: at the time of writing this paper, an independent \underline{mathematical} derivation of the $\mathsf{TMF}$ class $[(E_8)_2]$ of the $(E_8)_2$ theory was not available, at least to the knowledge of the author. 

Computations in TMF involve the intricate use of spectral sequences \cite{BrunerRognes,Douglas2014-cl}, and physics-inspired reasoning based on dualities and RG flows can provide hints when conventional mathematical techniques prove to be intractable. To determine the TMF class of a given 2d $\cN=(0,1)$ SQFT, one can try to find another representative in its deformation class, one for which the TMF class is already known or is more easily computable. Indeed, this was a key motivation spelled out in the \hyperref[sec:intro]{\textcolor{black}{Introduction}}. The 
 philosophy of exploiting dualities and RG flows to connect field theories in the space of 2d $\cN=(0,1)$ SQFTs can naturally be applied to other classes of models. (See \cite[Sec. 5]{Tachikawa:2024ucm} for some recent applications to ten-dimensional heterotic strings.) Another potential direction is the study of TMF classes of UV theories with multiple vacua splitting into distinct IR theories -- along the lines of \eqref{eq:sumuvir}, but now with not all isomorphic summands (see footnote \ref{foot:moregeneral}). Finally, an investigation of TMF classes of internal CFTs that arise in certain heterotic string compactifications to lower dimensions is work in progress. 

\appendix
\section{Theta Functions\label{app:ThetaFunc}}
We follow \cite{Alvarez-Gaume:1986rcs} with only mild notational changes. The general theta function is defined by
\eqa{
\vartheta\left[\begin{array}{c}\theta \\ \phi\end{array}\right](z|\tau) = \sum_{n \in \bZ} \exp\left[\sfi \pi(n+\theta)^2\tau + 2\pi \sfi (n+\theta)(z + \phi)\right] ~.
}
Setting $z = 0$, we obtain
\eqa{
& \vartheta\left[\begin{array}{c}\theta \\ \varphi \end{array}\right](\tau) := \vartheta\left[\begin{array}{c}\theta \\ \phi\end{array}\right](0|\tau)&&= \eta(\tau) e^{2\pi \sfi\theta\phi} q^{\frac{\theta^2}{2} - \frac{1}{24}}\prod_{n=1}^{\infty}(1 + q^{n + \theta - \frac{1}{2}}e^{2\pi \sfi \phi})(1 + q^{n-\theta-\frac{1}{2}}e^{-2\pi \sfi  \phi}) \nonumber\\
& &&= \sum_{n=-\infty}^{\infty}\exp\left[ \sfi \pi(n+\theta)^2\tau + 2\pi \sfi  (n + \theta)\phi \right] ~,
}
where $q = e^{2\pi \sfi  \tau}$, and $\eta$ is the Dedekind eta function,
\eqa{
&\eta(\tau) &&= q^{\frac{1}{24}}\prod_{n=1}^{\infty}(1-q^{n}) = q^{\frac{1}{24}}\sum_{n=-\infty}^{\infty}(-1)^{n} q^{\frac{3n^2-n}{2}}~.
}
In particular,
\eqa{
& \vartheta_{3}(\tau) &&:= \vartheta\left[\begin{array}{c}0 \\ 0 \end{array}\right](\tau) = \prod_{n=1}^{\infty}(1-q^n)(1 + q^{n-\frac{1}{2}})^2 = \sum_{n=-\infty}^{\infty}q^{\frac{1}{2}n^2} ~,\\
& \vartheta_{4}(\tau) &&:= \vartheta\left[\begin{array}{c}0 \\ \frac{1}{2} \end{array}\right](\tau) = \prod_{n=1}^{\infty}(1-q^n)(1 - q^{n-\frac{1}{2}})^2 = \sum_{n=-\infty}^{\infty}(-1)^{n}q^{\frac{1}{2}n^2}~,\\
& \vartheta_{2}(\tau) &&:= \vartheta\left[\begin{array}{c}\frac{1}{2} \\ 0 \end{array}\right](\tau) = 2 q^{1/8}\prod_{n=1}^{\infty}(1-q^n)(1 + q^n)^2  = \sum_{n=-\infty}^{\infty}q^{\frac{1}{2}\left(n + \frac{1}{2}\right)^2}~,\\
& \vartheta_{1}(\tau) &&:= \vartheta\left[\begin{array}{c}\frac{1}{2} \\ \frac{1}{2} \end{array}\right](\tau) = \sfi q^{1/8}\prod_{n=1}^{\infty}(1-q^n)^2(1 -q^{n-1}) = 0 ~, \label{eq:vartheta1-vanish}
}
They satisfy
\eqa{
& \vartheta_{2}^{4}(\tau) - \vartheta_{3}^{4}(\tau) + \vartheta_{4}^{4}(\tau) &&= 0 ~. \label{eq:obscure}
}
Further, under modular transformations,
\eqa{
&\vartheta\left[\begin{array}{c} \theta \\ \phi \end{array}\right](\tau + 1) &&= e^{-\sfi\pi(\theta^2-\theta)}\,\,\vartheta\left[\begin{array}{c} \theta \\ \theta+\phi-\frac{1}{2} \end{array}\right](\tau) ~,\\
&\vartheta\left[\begin{array}{c} \theta \\ \phi \end{array}\right]\left(-\frac{1}{\tau}\right) &&= (-\sfi\tau)^{\frac{1}{2}}\,\,\vartheta\left[\begin{array}{c} \phi \\ -\theta \end{array}\right](\tau) ~,\\
&\eta(\tau+1) &&= e^{\frac{\sfi\pi}{12}}\eta(\tau) ~, \\
&\eta\left(-\frac{1}{\tau}\right) &&= (-\sfi\tau)^{\frac{1}{2}}\eta(\tau) ~,
}
and under shifts of the characteristics,
\eqa{
& \vartheta\left[\begin{array}{c}\theta + a \\ \phi + b \end{array}\right](\tau) &&= e^{2\pi \sfi  \theta b}\vartheta\left[\begin{array}{c}\theta \\ \phi \end{array}\right](\tau) ~, \qquad a, b \in \bZ ~.
}
Therefore,
\eqa{
&\vartheta_{2}(\tau+1) &&= e^{\frac{\sfi\pi}{4}}\vartheta_{2}(\tau) ~, \qquad && \vartheta_{2}\left(-\frac{1}{\tau}\right) &&= (-\sfi\tau)^{1/2}\vartheta_{4}(\tau) ~, \\
&\vartheta_{3}(\tau+1) &&= \vartheta_{4}(\tau) ~, \qquad && \vartheta_{3}\left(-\frac{1}{\tau}\right) &&= (-\sfi\tau)^{1/2}\vartheta_{3}(\tau) ~,\\
&\vartheta_{4}(\tau+1) &&= \vartheta_{3}(\tau) ~, \qquad && \vartheta_{4}\left(-\frac{1}{\tau}\right) &&= (-\sfi\tau)^{1/2}\vartheta_{2}(\tau) ~.
}
It is also useful to note that
\eqa{
&\vartheta_1'(\tau) &&= 2\pi q^{\frac{1}{8}}\sum_{n=0}^{\infty}(-1)^{n}(2n+1)q^{\frac{n(n+1)}{2}} = 2\pi \eta^3(\tau) ~.
}

\section{Characters of $(E_8)_{1} \times (E_8)_{1}$ and $(E_8)_2$\label{app:compchar}}
The weight vectors of $E_8$ are
\eqa{
& \bm{\lambda} &&= \left\{ \begin{array}{l}(n_1, \ldots, n_8) ~,\\
\left(n_1 + \frac{1}{2}, \ldots, n_8 + \frac{1}{2}\right)\end{array} \right. \quad \sum_{i=1}^{8}n_{i} = \text{ even integer}. \label{eq:weights_e8}
}
The lattice sum for $E_8$,
\eqa{
& P_{E_8} &&= \sum_{\bm{\lambda}}e^{i\pi\tau\bm{\lambda}^2} ~,
}
can be rewritten by inserting $\frac{1+e^{i\pi\sum_{i=1}^{8}n_i}}{2}$ which enforces the $\sum_{i=1}^{8}n_{i} = \text{even}$ constraint in \eqref{eq:weights_e8}. This yields
\eqa{
& P_{E_8} &&= \frac{1}{2}\left[\prod_{i=1}^{8}\sum_{n_i \in \bZ}e^{i\pi n_{i}^2\tau} + \prod_{i=1}^{8}\sum_{n_i \in \bZ}e^{i\pi n_i^2\tau}e^{i\pi n_i} + \prod_{i=1}^{8}\sum_{n_i \in \bZ}e^{i\pi\left(n_i + \frac{1}{2}\right)^2\tau} + \prod_{i=1}^{8}\sum_{n_i \in \bZ}e^{i\pi\left(n_i + \frac{1}{2}\right)^2}e^{i\pi\left(n_i + \frac{1}{2}\right)} \right] \nonumber\\
& &&= \frac{1}{2}\left[\vartheta_{3}^{8}(\tau) + \vartheta_{4}^{8}(\tau) + \vartheta_{2}^{8}(\tau)\right] ~,
}
where, in the last line, we have expressed the result in terms of Jacobi $\vartheta$ functions reviewed in Appendix \ref{app:ThetaFunc}. For $(E_8 \times E_8)_{1}$, the relevant lattice sum is 
\eqa{
&P_{E_8}^2 &&= \frac{1}{4}\left[\vartheta_{3}^{8}(\tau) + \vartheta_{4}^{8}(\tau) + \vartheta_{2}^{8}(\tau)\right]^2 = 1 + 480q + 61920 q^2 + 1050240 q^3 + \mathcal{O}(q^4) ~.
}
Under modular transformations, $P_{E_8}^2(\tau+1) = P_{E_8}^2(\tau)$ and $P_{E_8}^2(-\frac{1}{\tau}) = \tau^8 P_{E_8}^2(\tau)$.

Following \cite{BoyleSmith:2023xkd} and \cite{Forgacs:1988iw}, we define the (holomorphic) Ising characters as
\eqa{
& \chi_{0}^{\Vir} &&= \sB_{I}^{0}(\tau) = \frac{\sqrt{\vartheta_{3}(\tau)} + \sqrt{\vartheta_{4}(\tau)}}{2\sqrt{\eta(\tau)}} = \frac{1}{2}q^{-1/48}\left\{ \prod_{n=1}^{\infty}\left(1 - q^{n-1/2}\right) + \prod_{n=1}^{\infty}\left(1 + q^{n-1/2}\right) \right\} ~,\\
& \chi_{1/2}^{\Vir} &&= \sB_{I}^{1}(\tau) = \frac{\sqrt{\vartheta_{3}(\tau)} - \sqrt{\vartheta_{4}(\tau)}}{2\sqrt{\eta(\tau)}} = \frac{1}{2}q^{-1/48}\left\{ \prod_{n=1}^{\infty}\left(1 - q^{n-1/2}\right) - \prod_{n=1}^{\infty}\left(1 + q^{n-1/2}\right) \right\} ~,\\
& \chi_{1/16}^{\Vir} &&= \sB_{I}^{2}(\tau) = q^{1/24}\prod_{n=1}^{\infty}(1 + q^{n}) = \frac{1}{\sqrt{2}}\sqrt{\frac{\vartheta_2(\tau)}{\eta(\tau)}} ~.
}
Under modular transformations, these transform as
\eqa{
& \mathcal{T}_{\mathsf{Ising}}\begin{pmatrix} \sB_{I}^{0}(\tau) \\ \sB_{I}^{1}(\tau) \\ \sB_{I}^{2}(\tau) \end{pmatrix} &&= \begin{pmatrix} e^{-i\pi/24} & 0 & 0 \\
0 & -e^{-i\pi/24} & 0 \\
0 & 0 & e^{i\pi/12}
\end{pmatrix} \begin{pmatrix} \sB_{I}^{0}(\tau) \\ \sB_{I}^{1}(\tau) \\ \sB_{I}^{2}(\tau) \end{pmatrix} 
~,\\
& \mathcal{S}_{\mathsf{Ising}}  \begin{pmatrix}\sB_{I}^{0}(\tau) \\ \sB_{I}^{1}(\tau) \\ \sB_{I}^{2}(\tau) \end{pmatrix} &&=
\begin{pmatrix} \frac{1}{2} & \frac{1}{2} & \frac{1}{\sqrt{2}} \\
\frac{1}{2} & \frac{1}{2} & -\frac{1}{\sqrt{2}} \\
\frac{1}{\sqrt{2}} & -\frac{1}{\sqrt{2}} & 0 
\end{pmatrix}\begin{pmatrix}\sB_{I}^{0}(\tau) \\ \sB_{I}^{1}(\tau) \\ \sB_{I}^{2}(\tau) \end{pmatrix} ~.
}
One can verify that $\mathcal{S}_{\mathsf{Ising}}^2 = 1$ and $(\mathcal{S}_{\mathsf{Ising}}\mathcal{T}_{\mathsf{Ising}})^3 = 1$.

The $(E_8)_{2}$ characters are defined as
\eqa{
& \chi_{\bm{1}}^{E_8} &&= \sB_{E_8}^{0}(\tau) = \frac{1}{\eta^{31/2}}\left[ \frac{1}{2}P_{E_8}^2\left(\frac{1}{\sqrt{\vartheta_4}} + \frac{1}{\sqrt{\vartheta_3}}\right) - \frac{1}{4}\left(\frac{1}{\sqrt{\vartheta_3}}P_3 + \frac{1}{\sqrt{\vartheta_4}}P_2 \right) \right] ~,\\
& \chi_{\bm{3875}}^{E_8} &&= \sB_{E_8}^{1}(\tau) = \frac{1}{\eta^{31/2}}\left[ \frac{1}{2}P_{E_8}^2\left(\frac{1}{\sqrt{\vartheta_3}} - \frac{1}{\sqrt{\vartheta_4}}\right) - \frac{1}{4}\left(\frac{1}{\sqrt{\vartheta_3}}P_3 - \frac{1}{\sqrt{\vartheta_4}}P_2 \right) \right] ~,\\
& \chi_{\bm{248}}^{E_8} &&= \sB_{E_8}^{2}(\tau) = \frac{\sqrt{2}}{\eta^{31/2}}\frac{1}{\sqrt{\vartheta_2}}\left[\frac{1}{2}P_{E_8}^2 - \frac{1}{4}P_1\right] ~, \label{eq:chi-e8level2-248}
}
where
\eqa{
&P_{1}(\tau) &&:= \vartheta_{3}^4 \vartheta_{4}^4 \left\{ \frac{1}{8}\left(\vartheta_3^8 + \vartheta_4^8\right) + \frac{7}{4}\vartheta_3^4 \vartheta_4^4\right\} ~,\\
&P_{2}(\tau) &&:= \vartheta_{3}^4 \vartheta_{2}^4 \left\{ \frac{1}{8}\left(\vartheta_3^8 + \vartheta_2^8\right) + \frac{7}{4}\vartheta_3^4 \vartheta_2^4\right\} ~,\\
& P_{3}(\tau) &&:= -\vartheta_{4}^4 \vartheta_{2}^4 \left\{ \frac{1}{8}\left(\vartheta_4^8 + \vartheta_2^8\right) - \frac{7}{4}\vartheta_4^4 \vartheta_2^4\right\} ~.
}
Under modular transformations, the $P_{i}$'s transform as follows:
\eqa{
& P_{1}(\tau+1) &&= P_1(\tau) ~, \qquad && P_{1}\left(-\frac{1}{\tau}\right) &&= \tau^8 P_2(\tau) ~,\\
& P_{2}(\tau+1) &&= P_3(\tau) ~, \qquad && P_{2}\left(-\frac{1}{\tau}\right) &&= \tau^8 P_1(\tau) ~,\\
& P_{3}(\tau+1) &&= P_2(\tau) ~, \qquad && P_{3}\left(-\frac{1}{\tau}\right) &&= \tau^8 P_3(\tau) ~.
}
Therefore, the modular transformations of the $(E_8)_2$ characters are
\eqa{
& \mathcal{T}_{E_8} \begin{pmatrix} \sB_{E_8}^{0} \\ \sB_{E_8}^{1} \\ \sB_{E_8}^{2} \end{pmatrix} &&= \begin{pmatrix} e^{-31i\pi/24} & 0 & 0 \\ 0 & -e^{-31i\pi/24} & 0 \\ 0 & 0 & e^{-17i\pi/12} \end{pmatrix} \begin{pmatrix} \sB_{E_8}^{0} \\ \sB_{E_8}^{1} \\ \sB_{E_8}^{2} \end{pmatrix} ~,\\
& \mathcal{S}_{E_8} \begin{pmatrix} \sB_{E_8}^{0} \\ \sB_{E_8}^{1} \\ \sB_{E_8}^{2} \end{pmatrix} &&= \begin{pmatrix} \frac{1}{2} & \frac{1}{2} & \frac{1}{\sqrt{2}} \\ \frac{1}{2} & \frac{1}{2} & -\frac{1}{\sqrt{2}} \\ \frac{1}{\sqrt{2}} & -\frac{1}{\sqrt{2}} & 0 \end{pmatrix}\begin{pmatrix} \sB_{E_8}^{0} \\ \sB_{E_8}^{1} \\ \sB_{E_8}^{2} \end{pmatrix} ~.
}
One can check that $\mathcal{S}_{E_8}^2 = 1$ and $(\mathcal{S}_{E_8}\mathcal{T}_{E_8})^3 = 1$. We note that $\mathcal{S}_{E_8} = \mathcal{S}_{\mathsf{Ising}}$. Furthermore, we note that $\mathcal{S}_{\mathsf{Ising},E_8}$ and $\mathcal{T}_{\mathsf{Ising},E_8}$ have determinant $-1$.

It is easy to verify the following identity:
\eqa{
& \eta^{-16} P_{E_8}^2 &&= \sB_{E_8}^{0}\sB_{I}^{0} + \sB_{E_8}^{1}\sB_{I}^{1} + \sB_{E_8}^{2}\sB_{I}^{2} ~. \label{eq:character-identity}
}
This implies that
\eqa{
& \chi_{\mathsf{0,0}} &&= \mathsf{tr }_{\mathsf{untwisted}}\,q^{L_0 - \frac{c}{24}} = \eta^{-16} P_{E_8}^2 = \sum_{i,j \in \{0,1,2\}} c_{ij} \sB_{E_8}^{i}\sB_{I}^{j}  ~, \qquad \text{ where } c_{ij} = \delta_{ij} ~.
}
The $\bZ_2$ action exchanges the two $(E_8)_1$ factors, but this action commutes with the $(E_8)_{2} \times \lambda$ decomposition, and therefore the coefficients $c_{ij}'$ in the resulting decomposition can only be $\pm 1$. A detailed analysis of the $\mathcal{H}_{\mathsf{s}}$ and $\mathcal{H}_{\mathsf{as}}$ subspaces of the Hilbert space \cite{Forgacs:1988iw} reveals that, in fact, $(c_{00}', c_{11}', c_{22}') = (0, 1, -1)$. Therefore,
\eqa{
& \chi_{\mathsf{0,1}} &&= \mathsf{tr }_{\mathsf{untwisted}}\,g\,q^{L_0 - \frac{c}{24}}   =  \sB_{E_8}^{0}\sB_{I}^{0} + \sB_{E_8}^{1}\sB_{I}^{1} - \sB_{E_8}^{2}\sB_{I}^{2} ~. 
}
The $\mathcal{S}$-transformation of $\chi_{\mathsf{0,1}}$ yields
\eqa{
& \chi_{\mathsf{1,0}} &&= \mathsf{tr }_{\mathsf{twisted}}\,q^{L_0 - \frac{c}{24}} =  \sB_{E_8}^{1} \sB_{I}^{0} + \sB_{E_8}^{0} \sB_{I}^{1} + \sB_{E_8}^{2} \sB_{I}^{2} ~,
}
and finally, the $\mathcal{T}$ transformation of $\chi_{\mathsf{1,0}}$ yields
\eqa{
& e^{-4\pi \sfi /3}\chi_{\mathsf{1,1}} &&= e^{-4\pi \sfi /3}\mathsf{tr }_{\mathsf{twisted}}\,g\,q^{L_0 - \frac{c}{24}} = -e^{2\pi \sfi /3}\left( \sB_{E_8}^{1} \sB_{I}^{0} + \sB_{E_8}^{0} \sB_{I}^{1} - \sB_{E_8}^{2} \sB_{I}^{2} \right) ~, \nonumber
}
which yields
\eqa{
& \chi_{\mathsf{1,1}} &&= \mathsf{tr }_{\mathsf{twisted}}\,g\,q^{L_0 - \frac{c}{24}} = -\sB_{E_8}^{1} \sB_{I}^{0} - \sB_{E_8}^{0} \sB_{I}^{1} + \sB_{E_8}^{2} \sB_{I}^{2} ~.
}

\section{Factorizing internal SCFT partition functions}\label{app:factoringintcft}

For a 2d CFT with central charge $(c_L, c_R)$, the Hilbert space decomposes in terms of Verma modules as
\eqa{
& \cH &&= \bigoplus_{\Delta,\ov{\Delta}}c_{\Delta,\ov{\Delta}} \cV(c_L, \Delta)  \otimes \cV(c_R, \Delta) ~,
}
where $\{c_{\Delta,\ov{\Delta}}\}$ is a theory-dependent set of coefficients. The partition function has the schematic form
\eqa{
& \sZ(\tau,\ov{\tau}) &&= \sum_{\Delta,\ov{\Delta}}c_{\Delta,\ov{\Delta}}\chi_{\Delta}(q)\chi_{\ov{\Delta}}(\ov{q}) ~. \label{eq:general-2d-cft-pf}
}
For unitary CFTs $c_{L } \geq 1$ and $c_{R} \geq 1$ (see \cite[Ch. 15]{Polchinski:1998rr} and \cite[Ch. 7]{DiFrancesco:1997nk}), and the only module inside $\cV(c_L, \Delta) \otimes \cV(c_R, \ov{\Delta})$ that has a null state is the vacuum module with $\Delta = 0$ and $\ov{\Delta} = 0$. 
So, all Verma modules other than the vacuum module are irreducible (that is, devoid of null submodules containing singular vectors or null states). 
 To summarize,
\eqa{
&\chi_{\Delta}(q) &&= \left\{ \begin{array}{ll}
   \frac{q^{\Delta + (1-c_L)/24}}{\eta(\tau)}  & \text{ for a non-vacuum module ($\Delta \neq 0$)} ~, \\
   \frac{(1-q)q^{(1-c_L)/24}}{\eta(\tau)} & \text{ for a vacuum module ($\Delta = 0$)}  ~,
\end{array} \right. \label{eq:VirasoroVaccumAndNonVacuum}
}
with a similar expression for $\chi_{\ov{\Delta}}(\ov{q})$ with $c_L \to c_R$, $q \to \ov{q}$, $\tau \to \ov{\tau}$, and $\Delta \to \ov{\Delta}$. Thus \eqref{eq:general-2d-cft-pf} has the form
\eqa{
\hspace{-0.22in}\sZ(\tau,\ov{\tau}) &= \frac{q^{\frac{(1-c_L)}{24}} \ov{q}^{\frac{(1-c_R)}{24}}}{|\eta(\tau)|^2}\bigg( c_{0,0}(1-q)(1-\ov{q}) + (1-q)\sum_{\ov{\Delta} \neq 0} c_{0,\ov{\Delta}} \ov{q}^{\ov{\Delta}} + (1-\ov{q})\sum_{\Delta \neq 0}c_{\Delta,0}q^{\Delta}  + \sum_{\substack{\Delta \neq 0 \\ \ov{\Delta} \neq 0}} c_{\Delta,\ov{\Delta}} q^{\Delta}\ov{q}^{\ov{\Delta}} \bigg)
 ~. \label{eq:general-2d-cft-pf-final}
}
Next, consider a generic 2d $\cN = (0,1)$ SCFT with central charge $(c_L, c_R)$. As we will admit theories for which $c_L - c_R$ could be a half-integer, we restrict our attention to the $\sN\sS$ sector, where we have the following well-defined left-moving chiral characters:
\eqa{
& \chi_{\sN\sS,\sN\sS} &&= \mathsf{Tr}_{\sN\sS}[ q^{L_0 - \frac{c_L}{24}} ] = q^{\Delta - \frac{c_L}{24}}\prod_{n=1}^{\infty}\frac{1 + q^{n-\frac{1}{2}}}{1 - q^n} ~, \label{eq:nschar}\\
& \chi_{\sN\sS,\sR} &&= \mathsf{Tr}_{\sN\sS}[ (-1)^{\sF}q^{L_0 - \frac{c_L}{24}} ] = q^{\Delta - \frac{c_L}{24}}\prod_{n=1}^{\infty}\frac{1 - q^{n-\frac{1}{2}}}{1 - q^n} ~. \label{eq:nschar2}
}
But we must be careful about the vacuum module as in the bosonic case. First of all, note that for the $\sN\sS$ sector, $L_{-1} = G_{-1/2}^2$ (here $\{G_{r}\}_{r\in \mathbb{Z} + \frac{1}{2}}$ denote the modes of the supercurrent, and $\{L_{m}\}_{m\in\IZ}$ denote the modes of the energy-momentum tensor). So, since $L_{-1}$ annihilates the vacuum state, so does $G_{-1/2}$. Therefore, the characters \eqref{eq:nschar} and  \eqref{eq:nschar2} should be refined to
\eqa{
& \chi_{\sN\sS,\sN\sS} &&= \left\{
\begin{array}{ll}
 q^{\Delta - \frac{c_L}{24}}\prod_{n=1}^{\infty}\frac{1 + q^{n-\frac{1}{2}}}{1 - q^n} ~, & \text{ for a non-vacuum module ($\Delta \neq 0$)} ~,\\
 q^{-\frac{c_L}{24}} ~, \prod_{n=2}^{\infty}\frac{1 + q^{n-\frac{1}{2}}}{1 - q^n} & \text{ for the vacuum module ($\Delta = 0$)} ~.
 \end{array}
\right.\label{eq:c5}\\
  & \chi_{\sN\sS,\sR} &&= \left\{
    \begin{array}{ll}
     q^{\Delta - \frac{c_L}{24}}\prod_{n=1}^{\infty}\frac{1 - q^{n-\frac{1}{2}}}{1 - q^n} ~, & \text{ for a non-vacuum module ($\Delta \neq 0$)} ~,\\
     q^{-\frac{c_L}{24}} \prod_{n=2}^{\infty}\frac{1 - q^{n-\frac{1}{2}}}{1 - q^n} ~, & \text{ for the vacuum module ($\Delta = 0$)} ~.
     \end{array}
     \right. \label{eq:c6}
}
Let us define
\eqa{
&f_{+}(q) := \prod_{n=1}^{\infty} \frac{1+q^{n-\frac{1}{2}}}{1-q^n} &&= 1 + q^{1/2} + q + 2 q^{3/2} + 3 q^2 + 4 q^{5/2} + 5 q^3 + \cdots = \sum_{n \in \mathbb{Z}_{\geq 0}} p_{n} q^{n/2}~, \label{eq:fplus}\\
&f_{-}(q) := \prod_{n=1}^{\infty} \frac{1-q^{n-\frac{1}{2}}}{1-q^n} &&= 1 - q^{1/2} + q - 2 q^{3/2} + 3 q^2 - 4 q^{5/2} + 5 q^3 - \cdots = \sum_{n \in \mathbb{Z}_{\geq 0}} (-1)^{n} p_{n} q^{n/2} ~. \label{eq:fminus}
}
Note that
\eqa{
& \prod_{n=2}^{\infty} \frac{1 \pm q^{n-\frac{1}{2}}}{1-q^{n}} &&= (1 \mp q^{1/2})\prod_{n=1}^{\infty} \frac{1 \pm q^{n-\frac{1}{2}}}{1-q^{n}} = (1 \mp q^{1/2})f_{\pm}(q) ~.
}
The partition function of an internal 2d $\cN=(0,1)$ SCFT in the $(\sN\sS,\sN\sS)$ sector is
\eqa{
  &\sZ_{\sN\sS,\sN\sS} &&= \sum_{Δ,\ov{Δ}}C_{Δ,\ov{Δ}} \chi_{\sN\sS,\sN\sS,Δ}(q) \chi_{\sN\sS,\sN\sS,\ov{Δ}}(\ov{q}) 
  =  q^{-c_L/24} \ov{q}^{\,-c_R/24} |f_{+}(q)|^2 \sum_{Δ,\ov{Δ}} \alpha_{\ov{Δ},Δ} q^{Δ} \ov{q}^{\ov{Δ}} ~, \label{eq:Z (0,1) NSNS} 
}
where $\alpha_{\ov{Δ},Δ}$ is related to $C_{\Delta,\ov{\Delta}}$ by
\eqas{
\alpha_{0,0} &= C_{0,0} ~, \quad 
\alpha_{\frac{1}{2},0} = -C_{0,0} + C_{\frac{1}{2},0} ~, \quad
\alpha_{0,\frac{1}{2}} = -C_{0,0} + C_{0,\frac{1}{2}} ~, \quad
\alpha_{\frac{1}{2},\frac{1}{2}} = C_{0,0} + C_{\frac{1}{2},\frac{1}{2}} ~,\\
\alpha_{0,\ov{Δ}} &= C_{0,\ov{Δ}} ~, \quad \alpha_{\frac{1}{2},\ov{Δ}} =-C_{0,\ov{Δ}}  \,\,\,\,\,\,\text{ if } \quad  \ov{Δ} \neq 0 ~,\\
\alpha_{Δ,0} &= C_{Δ,0} ~, \quad 
\alpha_{Δ,\frac{1}{2}} = -C_{Δ,0} \quad \text{ if } \quad Δ \neq 0 ~,\\
\alpha_{Δ,\ov{Δ}} &= C_{Δ,\ov{Δ}} \quad\,\,\, \text{ if } \quad  Δ \neq 0, \frac{1}{2} \text{\,\,and\,\,} \ov{Δ} \neq 0, \frac{1}{2} ~.
}
Under a $\cT$-transformation ($\tau \mapsto \tau+1$ and $\ov{\tau} \mapsto \ov{\tau} + 1$), $f_{+}(q) \mapsto f_{-}(q)$.
We define
\eqa{
  & \sZ_{\sN\sS,\sR} &&=  q^{-c_L/24} \ov{q}^{\,-c_R/24} e^{-2\pi \sfi  \frac{(c_L-c_R)}{24}}|f_{-}(q)|^2 \sum_{Δ,\ov{Δ}} e^{2\pi \sfi (Δ-\ov{Δ})}\alpha_{\ov{Δ},Δ} q^{Δ} \ov{q}^{\ov{Δ}} ~. \label{eq:Z (0,1) NSR}
}
Note that $(Δ + \frac{n}{2}, \ov{Δ} + \frac{m}{2})$ is the conformal dimension of the state at level $(n,m)$ in the conformal family $(Δ, \ov{Δ})$. Therefore, defining $\Delta + \frac{n}{2} := \Delta'$ and $\ov{\Delta} + \frac{m}{2} := \ov{\Delta}'$, we can recast the partition functions as
\eqa{
  & \sZ_{\sN\sS,\sN\sS} &&= q^{-c_L/24} \ov{q}^{\,-c_R/24}\sum_{Δ', \ov{Δ}'} N_{Δ',\ov{Δ}'} q^{Δ'}\ov{q}^{\ov{Δ}'} ~, \label{eq:Z (0,1) NSNS refined} \\
  & \sZ_{\sN\sS,\sR} &&= q^{-c_L/24}  \ov{q}^{\,-c_R/24} e^{-2\pi \sfi \frac{(c_L-c_R)}{24}} \sum_{Δ', \ov{Δ}'} N_{Δ',\ov{Δ}'} e^{2\pi \sfi (Δ'-\ov{Δ}')} q^{Δ'}\ov{q}^{\ov{Δ}'} ~, \label{eq:Z (0,1) NSR refined}
}
where
\eqa{
  & N_{Δ',\ov{Δ}'} &&:= \sum_{n,m \in \mathbb{Z}_{\geq 0}} \alpha_{\ov{Δ}'-\frac{m}{2},Δ'-\frac{n}{2}}p_{n}p_{m} ~.
}

\interlinepenalty=10000
\addcontentsline{toc}{section}{References} 
\bibliographystyle{ytphys}  
\bibliography{refse8paper} 

\end{document}